\documentclass[10 pt]{article}
\usepackage[latin1]{inputenc}
\usepackage{amsmath}

\usepackage{amsthm}

\usepackage{hyperref}
\usepackage{braket}
\usepackage{amsfonts}
\usepackage{amsbsy}
\usepackage{amssymb}
\usepackage{amsmath}
\usepackage{bm}
\usepackage{bbold}
\usepackage{amscd}
\usepackage{makeidx}
\usepackage[parfill]{parskip}
\usepackage{pb-diagram}
\usepackage[all]{xy}
\usepackage{graphicx}
\usepackage{subcaption}
\usepackage{float}
\usepackage{tikz}
\usetikzlibrary{decorations.markings,decorations.pathmorphing,matrix}
\tikzset{->-/.style={decoration={
  markings,
  mark=at position #1 with {\arrow[arrowstyle]{stealth}}},postaction={decorate}},
particle/.style={postaction={decorate}, decoration={markings,mark=at position .5 with {\arrow{>}}}},
antiparticle/.style={postaction={decorate}, decoration={markings,mark=at position .5 with {\arrow{<}}}},
graviton/.style={decorate, decoration={coil,amplitude=4pt, segment length=5pt}}}

\numberwithin{equation}{section}

\newcommand{\rf}[1]{{(\ref{#1})}}
\newcommand{\appropto}{\mathrel{\vcenter{
  \offinterlineskip\halign{\hfil$##$\cr
    \propto\cr\noalign{\kern2pt}\sim\cr\noalign{\kern-2pt}}}}}

\makeatletter
\newcommand*{\rom}[1]{\expandafter\@slowromancap\romannumeral #1@}
\makeatother

\begin{document}

\begin{titlepage}
\begin{flushright}
\end{flushright}

\vspace{20pt}

\begin{center}

{\Large\bf Geometric flux formula for the gravitational Wilson loop
}
\vspace{15pt}

{\large N.\ Klitgaard$^{a, \sharp}$, R.\ Loll$^{a, \flat}$, M.\ Reitz$^{a, \dag}$, R.\ Toriumi$^{b,\ddag} $}

\vspace{15pt}

$^{a}${\sl Institute for Mathematics, Astrophysics, and Particle Physics, Radboud University, }\\
{\sl Heyendaalseweg 135, 6525 AJ Nijmegen, The Netherlands }

\vspace{5pt}

$^{b}${\sl Okinawa Institute of Science and Technology Graduate University\\ 
{\sl 1919-1 Tancha, Onna, Kunigami, Okinawa, Japan 904-0412}}\\

\vspace{15pt}

emails:  {\sl 
$^{\sharp}$n.klitgaard@science.ru.nl,
$^{\flat}$r.loll@science.ru.nl,
$^{\dag}$m.reitz@science.ru.nl,
$^\ddag$reiko.toriumi@oist.jp}

\vspace{40pt}

\begin{abstract}
\noindent Finding diffeomorphism-invariant observables to characterize the properties of gravity and spacetime at the Planck scale 
is essential for making progress in quantum gravity. The holonomy and Wilson loop of the Levi-Civita connection are 
potentially interesting ingredients in the construction of quantum curvature observables. 
Motivated by recent deve\-lopments in nonperturbative quantum gravity, we establish new relations in three and four dimensions 
between the holonomy of a finite loop and certain curvature integrals over the surface spanned by the loop.
They are much simpler than a gravitational version of the nonabelian Stokes' theorem, 
but require the presence of totally geodesic surfaces in the manifold, which follows from the existence of suitable Killing vectors.   
We show that the relations are invariant under smooth surface deformations, due to the presence of a conserved geometric flux.

\end{abstract}

\vspace{10pt}

\today
\end{center}


\bigskip

\setcounter{footnote}{0}

\end{titlepage}

\tableofcontents
\newpage

\section{Introduction}
\label{sec:intro}

The motivation and larger context for the work presented here is the search for observables in nonperturbative quantum gravity. 
Such observables are key to understanding the dynamics of gravity and the nature of spacetime at the Planck scale -- whichever form these
concepts will take in a nonperturbative regime -- and crucial for determining whether a given candidate theory has the correct classical limit. 
More specifically, we are interested in quantum observables relating to curvature, a subject about which currently little is known. 
One strand of research that is being explored systematically in the context of Causal Dynamical Triangulations (CDT), a nonperturbative
lattice approach to quantum gravity \cite{review1,review2}, 
is the implementation and measurement of quantum Ricci curvature \cite{qrc1,qrc2,qrc3}.
The work presented below uses a different ansatz and takes place on more familiar territory. 
It investigates the question of whether gravitational holonomies of the Levi-Civita connection along
non-infinitesimal closed curves and their associated Wilson loops can be used to construct observables
from which one can retrieve curvature information about the underlying space, in the spirit of a (perhaps generalized) Stokes' theorem. 

A natural first step is to examine the answer to this question in classical gravity, to guide one's intuition in the 
quantum theory and to understand which classical limit should be obeyed by the quantum construction. However, 
for classical curved manifolds
no relevant and useful results seem to be known on how to relate the holonomy of finite loops to some form of integrated
or averaged curvature. The reason why one may expect such a relation to exist in the first place is the presence of the corresponding property for
infinitesimal loops. This is most familiar in the context of gauge field theory, where instead of the Levi-Civita connection $\Gamma$
one works with a gauge connection $A$. In this case, the holonomy (or path-ordered exponential) $W_{\gamma_{[\mu\nu ]},p}$
of $A$ along an infinitesimal square loop $\gamma_{[\mu\nu ]}$ in the $(\mu ,\nu)$-plane with side length $\varepsilon$ 
and base point $p$ can be expanded in powers of $\varepsilon$, leading to the well-known expression
\begin{equation}
W_{\gamma_{[\mu\nu ]},p} =\mathrm{P} \exp \oint_{\gamma_{[\mu\nu ]}} \!\!\!\!\! A=
\mathbb{1}+ 
\varepsilon^2 F_{\mu\nu}^a(p) X_a +o(\varepsilon^2),
\label{gaugewils}
\end{equation}
where the $X_a$ are the generators of the gauge Lie algebra, usually given by $su(N)$, and P indicates path-ordering. The important point is the
appearance in eq.\ (\ref{gaugewils}) of the $(\mu,\nu)$-component of the field strength tensor $F$ of the connection $A$, which can be read off 
directly from the lowest nontrivial order in the $\varepsilon$-expansion. As we will describe in detail in Sec.\ \ref{nasty}
below, an analogous relation holds for the holonomy of the connection $\Gamma$ 
of an infinitesimal square loop on a Riemannian manifold, with $F_{\mu\nu}$
replaced by the corresponding components of the Riemann curvature tensor. In other words, in either case there is a straightforward
relation between holonomy and curvature at the perturbative level. 

In both gauge theory and gravity, the key obstacle to extending (\ref{gaugewils}) to a similarly straightforward relation between holonomy 
and curvature for {\it non}-infinitesimal loops is the nonabelian nature of the underlying connection form. More precisely, there exists a relation
of this kind, which in a
gauge-theoretic context usually goes by the name of ``nonabelian Stokes' theorem" \cite{aref} and whose construction we review in Sec.\ \ref{nasty}. 
However, it is not particularly useful for our purposes because of its unwieldy, nonlocal functional form. The main source of complication is the
surface-ordering for the area integral appearing in the theorem, which is needed because of the noncommuting nature of the 
connection and associated field strength or curvature. 

The reason why we are primarily
interested in loops of finite size comes again from the quantum theory. Note that derivations in the continuum like that of eq.\ (\ref{gaugewils}) 
make crucial use of the smooth structure of the underlying geometry, which in a Planckian regime will typically not be present.  
This is certainly true for candidate theories of quantum gravity that postulate fundamental discreteness at the Planck scale, but related
issues also arise in other nonperturbative formulations.  
Let us turn again to the framework of CDT for illustration. Its regularized path integral (i.e. before taking any scaling limit) 
is based on ensembles of piecewise flat geometries, 
where the concept of an infinitesimal loop is not particularly meaningful or interesting. In this setting, one could in principle study
the smallest loops that have a nontrivial holonomy\footnote{These are loops that wind around a single curvature singularity located at a subsimplex of
codimension 2, see e.g. \cite{review1} for technical details.}, but they are 
of the size of the ultraviolet cutoff (the typical edge length or lattice spacing) and therefore dominated by
regularisation-dependent lattice artefacts, rather than containing interesting physical information. If one is interested in observables 
involving loop holonomies or their traces, the Wilson loops, one therefore has to deal with loops of finite size (in terms of lattice units), 
which are large compared to the lattice cutoff. At the same time, the geometry on these scales is far from flat, which means
that generic holonomies will {\it not} be of the form ``unit matrix plus a small perturbation", in contrast to eq.\ (\ref{gaugewils}). 
The example of CDT quantum gravity is particularly relevant, since -- unlike on general curved smooth manifolds -- the evaluation of
arbitrary holonomies is computationally straightforward \cite{wilsoncdt}. Therefore, {\it if} one was able to derive a sufficiently simple
relation between holonomies and curvature for non-infinitesimal loops, these could potentially be used in the construction of genuine curvature
observables on various scales, depending on the size of the underlying loops.

An important difference between gauge-theoretic and gravitational Wilson loops is that the former 
can be directly promoted to quantum observables, at least formally. Famously, the expectation value of the Wilson loop serves as
an order parameter for confinement in QCD \cite{wilsonconf,ukawa}. The situation in nonperturbative quantum gravity 
is more involved: the ``expectation value of the Wilson loop of a given loop $\gamma$" is not a meaningful concept, 
because it is not possible to identify one and the same loop $\gamma$ across the different spacetime configurations that
make up the quantum ensemble in which the expectation value is computed. This is why above we
have talked about Wilson loops and holonomies of the Levi-Civita connection only as possible ingredients in the construction of quantum observables, and not as observables in themselves, despite their classical invariance properties.  

Additional work is
required in the quantum theory to construct genuine observables from them. In a matter-coupled theory, one could mark the location of the
loop in terms of the matter present, before performing an ensemble average over geometries. Such a prescription was followed in
\cite{wilsoncdt}, which studied the expectation value of a Wilson loop whose underlying loop coincides with the worldline of a particle, 
cyclically identified in time.   
Note that this construction involved very long, noncontractible loops, whereas in the present work we are interested in contractible, 
nonintersecting loops of all sizes that
run along the boundary of a two-dimensional disc. One way how these may be turned into well-defined quantum observables in pure gravity would be
by averaging over subsets of loops with some specified geometric properties (e.g. a fixed value of their length and other invariant parameters describing 
their shape and size), before performing the path integral over geometries. While these are interesting and nontrivial issues, the present work will
focus exclusively on an analysis of the classical case, and the search for a relation between the gravitational holonomies and Wilson loops of finite-sized 
closed curves and the curvature of the underlying manifold. Our investigation will be conducted on Riemannian manifolds
$({\cal M},g_{\mu\nu})$ with a positive definite metric $g_{\mu\nu}$, which is the relevant framework for quantum gravity formulations with 
a Wick rotation or some other form of analytic continuation from Lorentzian signature, like CDT.  

Elsewhere in gravity, Wilson loops have appeared in a variety of contexts and with different motivations. An obvious area of application
are gauge-theoretic formulations of gravity. In the Chern-Simons formulation of three-dimensional gravity, noncontractible Wilson loops are used to
capture its (global) degrees of freedom in a gauge-invariant manner \cite{witten}. Loop quantum gravity derives its name from Wilson loop
variables defined on slices of constant time in spacetime, whose Poisson algebra served as a starting point for a nonperturbative 
canonical quantization in the original version of the theory \cite{rovsmo}. A similar type of canonical loop representation can also be 
constructed in three spacetime dimensions \cite{ashloll}.
In the standard metric formulation of gravity based on four-metrics $g_{\mu\nu}$, properties of the Wilson loop in perturbative quantum 
gravity on a Minkowskian background were investigated in \cite{modanese}.
In the context of quantum Regge calculus, another lattice approach to quantum gravity, an attempt was made to treat large Wilson loops 
in an almost-flat setting, and to perform a strong-coupling analysis of the gravitational Wilson loop along the lines of what is done in QCD \cite{hawi}. 

We are taking a different perspective here by asking whether and how Wilson loops may be useful in constructing {\it observables}
in nonperturbative quantum gravity, beyond a regime where fields are sufficiently weak and/or loops sufficiently small to work with 
perturbative expressions like eq.\ (\ref{gaugewils}). As outlined above, this has motivated our analysis of non-infinitesimal loop holonomies 
on classical Riemannian manifolds in dimensions three and four, and their relation to curvature. Not unexpectedly, given the complicated
and nonlocal functional form of the holonomy, we have not
been able to derive a simple relation between holonomy and curvature for {\it general} metrics and loops.\footnote{Since curvature is a tensorial
quantity, this might have given us new insights into the notorious ``averaging problem" of how to average tensors on a 
Riemannian manifold in a covariant way, see \cite{colellis} for a recent assessment 
of the ramifications of this issue for general relativity and cosmology.} Instead, we have derived a new relation of this kind in
a more restrictive setting, where the manifolds have symmetries that allow for the presence of so-called totally geodesic surfaces. 
While such manifolds are not generic, there are many examples of Riemannian spaces with Killing vectors that satisfy the
required technical conditions. We will show that the invariant angle(s) characterizing the holonomy of a loop $\gamma$ lying in one of the 
totally geodesic surfaces 
of such a manifold are directly related to ordinary two-dimensional curvature integrals over a disc bounded by $\gamma$. 

The remainder of this paper is structured as follows. In Sec.\ \ref{wilsonloopgrav}, we recall some details of the construction and properties of the holonomy of
the Levi-Civita connection on a Riemannian manifold, including the important concept of path ordering. This allows us to relate the holonomy of
an infinitesimal loop to the local Riemann tensor in Sec.\ \ref{nasty}, and to rederive the so-called nonabelian Stokes' theorem, which expresses the holonomy
of a finite loop in terms of a surface-ordered area integral depending on the curvature in a nonlocal way, a construction that goes back almost a hundred years.   
Sec.\ \ref{tgsurfaces} contains the core of our work. We demonstrate that a large class of three- and four-dimensional Riemannian manifolds 
with isometries possess non-infinitesimal loops whose holonomy is abelian and can be expressed in terms of standard surface integrals of suitable 
curvature scalars.
The holonomy group of the underlying manifold $\cal M$ need not be abelian, but the loops must lie in a leaf of a foliation of $\cal M$ by 
a family of totally geodesic surfaces.
We show in Sec.\ \ref{sec:d3} how the invariant angle characterizing the holonomy of such a loop in three dimensions
can be expressed as a surface integral, built up
from infinitesimal area contributions. The explicit computation of the surface integrals in three and four dimensions is performed in Secs.\ \ref{sec:threedim}
and \ref{sec:4d} respectively. For illustration, we apply the construction to a specific curved manifold, the round three-sphere, in Sec.\ \ref{sec:S3}. 
Interestingly, it turns out that the surface integrals in both three and four dimensions
are invariant under smooth surface deformations that leave the boundary loop
invariant. We show that this property is related to the existence of a conserved ``geometric flux" constructed from the Killing vector(s) and the Riemann tensor
of $\cal M$. The final Sec.\ \ref{sec:conclusions} contains a summary and a discussion of possible applications of our results.

\section{Holonomies and Wilson loops in gravity}
\label{wilsonloopgrav}

Given a $d$-dimensional manifold $\mathcal{M}$ with $d\geq 2$ and metric $g_{\mu\nu}$, we will be interested in the holonomy $U_\gamma[\Gamma]$,
depending on the metric-compatible Levi-Civita connection $\Gamma_{\mu  \lambda}^{\kappa}$ associated with $g_{\mu\nu}$,
and on a parametrized path
$\gamma: I\rightarrow \mathcal{M},\; \tau\mapsto \gamma (\tau)$, where $I$ denotes an interval $I=[\tau_0,\tau_1]$ on the real
line. We will deal 
with the Riemannian case, corresponding to either Euclidean gravity or ``gravity after a Wick rotation", as is the case in CDT, say.

The holonomy $U_\gamma (\tau,\tau_0)$ is the solution to the differential equation
\begin{equation}
\frac{d}{d\tau} U_\gamma (\tau,\tau_0) = -\Gamma_\mu (\gamma(\tau))\ \frac{d\gamma^\mu (\tau)}{d\tau}\ U_\gamma (\tau,\tau_0),\;\;\;
\tau_0\leq\tau\leq \tau_1, 
\label{holoeq}
\end{equation}
subject to the initial condition $U_\gamma (\tau_0,\tau_0)=\mathbb{1}$, the unit matrix.
Note that (\ref{holoeq}) is a matrix equation, with $(\Gamma_\kappa)^\mu{}_\nu :=\Gamma_{\kappa\nu}^\mu$. The holonomy takes values
in $GL(d,\mathbb{R})$ and describes how a vector $v$ at some initial point $x_0=\gamma (\tau_0)\in \mathcal{M}$ behaves under parallel 
transport along the curve $\gamma$ to some final point $x_1=\gamma(\tau_1)$, namely, according to the linear transformation
\begin{equation}
v^\mu(x_1)=\big( U_\gamma (\tau_1,\tau_0)\big)^\mu{}_\nu\, v^\nu(x_0).
\label{ptrans}
\end{equation}
It follows from the parallel-transport property that the inverse path $\gamma^{-1}$ is associated with a holonomy
that is the matrix inverse $U_\gamma^{-1}$ of $U_\gamma$ (see, for example, \cite{Loll:1993ht}).
Difficulties in computing the holonomy explicitly for a given connection $\Gamma$ and curve $\gamma$ come from 
the fact that the contraction $A(\tau)$ of the connection with the tangent vector $\dot{\gamma}=d\gamma/d\tau$ to the curve,
\begin{equation}
A^\mu{}_\nu (\tau):=-\Gamma_{\kappa\nu}^\mu (\gamma(\tau))\dot{\gamma}^\kappa(\tau),
\label{gammacontr}
\end{equation}
takes values in the nonabelian Lie algebra $gl(d,\mathbb{R})$, where two fields $A(\tau)$ for different values of $\tau$ will in general
not commute. In other words, we have to keep track of the factor order when integrating $A(\tau)$ along a path $\gamma$ to
obtain the holonomy $U_\gamma$. This explains the occurrence of the path-ordering symbol ``$\mathrm{P}$" in the standard notation for the
holonomy,
\begin{equation}
(U_\gamma (\tau_1,\tau_0))^\mu{}_\nu=\big({\rm \mathrm{P}\,  e}^{-\int_{\tau_0}^{\tau_1} d \tau \, \dot{\gamma}^{\kappa}(\tau) 
\Gamma_{\kappa}(\tau)}\big)^\mu{}_\nu
\equiv 
\big( {\rm \mathrm{P}\, e\, }^{\int_{\tau_0}^{\tau_1} d \tau \, A(\tau) } \big)^\mu{}_\nu ,
\label{wline}
\end{equation}
also called the {\it path-ordered exponential} of the Levi-Civita connection $\Gamma$ along the path $\gamma$.
The right-hand side of (\ref{wline}) can be defined as an infinite sum of nested integrals,
\begin{equation}
{\rm \mathrm{P}\, e\, }^{\int_{\tau_0}^{\tau_1} d \tau \, A(\tau) }  := \mathbb{1}+\sum_{n=1}^\infty \
\int_{\tau_0}^{\tau_1}\!\! dt_1 \int_{t_1}^{\tau_1}\!\! dt_2\, \dots \int_{t_{n-1}}^{\tau_1}\!\!\!\! dt_n\, A(t_n)A(t_{n-1})\dots A(t_1),
\label{holoexpand}
\end{equation}
where the factors of $A(t_i)$ in the integrand of the $n$th term in the sum are path-ordered from right to left since 
$\tau_0\leq t_1\leq t_2\leq \dots \leq t_{n-1} \leq t_n \leq \tau_1$ is enforced by the integration limits. 
An alternative way of defining the holonomy employs
a limiting process with ever finer finite approximations of the path $\gamma$,
\begin{equation}
U_\gamma (\tau_1,\tau_0):= \lim_{n\rightarrow\infty} (\mathbb{1}+A(t_n)\Delta_n)(\mathbb{1}+A(t_{n-1})\Delta_{n-1})\dots
(\mathbb{1}+A(t_1)\Delta_1),
\label{holoalt}
\end{equation}
where $\Delta_i$ is defined as $\Delta_i=t_i-t_{i-1}$, 
the parameters are again arranged in increasing order, $\tau_0 < t_1 <t_2<\dots <t_n=\tau_1$,
and $\sup \Delta_i\rightarrow 0$ as $n\rightarrow\infty$. The functional form (\ref{holoalt}) is that of a so-called 
product integral, of the type first introduced by the mathematician Volterra in the late 1800s.\footnote{see \cite{slavik} for a detailed 
account of the historical development of this notion} Yet another way of expressing the holonomy (see, for example, \cite{chafom}) 
is as the limit
\begin{equation}
U_\gamma =\lim_{n\rightarrow\infty} {\rm e}^{\, A(t_n)\Delta_n} {\rm e}^{\, A(t_{n-1})\Delta_{n-1}}\dots {\rm e}^{\, A(t_{1})\Delta_{1}}.
\label{holoother}
\end{equation}
From this variant of (\ref{holoalt}) it is straightforward to read off how $U(\gamma)$ simplifies if for some reason the 
fields $A(t_i)$ all commute with each other. In this case, one has 
\begin{equation}
U_\gamma =\lim_{n\rightarrow\infty} {\rm e}^{\,\sum_i\! A(t_i)\Delta_i} = {\rm e}^{\,\int dt\, A(t)}, \;\;\;\;\;\; {\rm (abelian\; case)}
\label{holoabel}
\end{equation}
by virtue of the Baker-Campbell-Hausdorff formula, resulting in the exponentiation of an ordinary integral of $A(t)$. 

As already mentioned in the introduction, path-ordered integrals of gauge connections have been used in
$SU(N)$-gauge field theory, where the analogue of the field $A$ of eq.\ (\ref{gammacontr}) is
$su(N)$-valued, and is obtained by contracting the gauge potential of the theory with the tangent vector to the curve $\gamma$. 
One attractive feature of using holonomies in gauge theory is the fact that one can use them to construct
gauge-invariant -- albeit nonlocal -- observables in terms of so-called Wilson loops, given by
traces of holonomies of closed curves (loops). 

Coming back to the gravitational case, to distinguish the holonomy of a closed curve $\gamma$, with a single base point
$x_0=\gamma(\tau_0)=\gamma(\tau_1)$, 
from that of a general curve given
in eq.\ (\ref{wline}) above, we will use the notation $W$ instead of $U$, that is,
\begin{equation}
(W_{\gamma,x_0})^\mu{}_\nu=\big({\rm \mathrm{P} \,  e}^{-\oint_\gamma d \tau \, \dot{\gamma}^{\kappa}(\tau) 
\Gamma_{\kappa}(\tau)}\big)^\mu{}_\nu .
\label{wloop}
\end{equation}
Since this holonomy describes the parallel transport of a vector $v\in T_{x_0}{\mathcal M}$ around $\gamma$, ending up in
the same tangent space, and since the parallel transport preserves the vector's norm, its effect on the vector is that
of an $SO(d)$-rotation.\footnote{Here and in what follows, we only consider contractible loops $\gamma$. In other words, we work
with the so-called restricted holonomy group.} 
If the basis of tangent space is chosen orthonormal, any holonomy $(W_\gamma)^\mu{}_\nu$ is of the form of a 
$d\times d$ rotation matrix in the defining representation of $SO(d)$. 

Irrespective of any choice of basis, if two closed paths $\gamma_1$ and $\gamma_2$ share a common base point, 
we can compose them into a single loop $\gamma_2\circ\gamma_1$ with respect to a suitably chosen common parameter
$\tau$, which first runs through $\gamma_1$ and then $\gamma_2$. The corresponding holonomy is obtained
by simply multiplying the matrices of the individual loop holonomies,
\begin{equation}
W_{\gamma_2\circ\gamma_1}=W_{\gamma_2}W_{\gamma_1},
\label{loopcomp}
\end{equation}
where in terms of notation we have suppressed the explicit reference to the common base point. 

As discussed in detail in \cite{wilsoncdt}, under a diffeomorphism of $\cal M$ the components of the gravitational holonomy 
$U_\gamma (\tau_1,\tau_0)$ transform
nontrivially at its two endpoints. This is still true when considering a loop holonomy $W_{\gamma,x_0}$ with base point $x_0$, but
taking the matrix trace Tr$\, W_{\gamma,x_0}$ removes this part of the diffeomorphism dependence 
(as well as the dependence on the base point $x_0$) 
because of the cyclicity of the trace.
The resulting quantity is the {\it gravitational Wilson loop}, schematically,
\begin{equation}
\mathrm{Tr}\, W_\gamma= \mathrm{Tr}\, {\rm \mathrm{P} \,  e}^{-\oint_\gamma \Gamma}.
\label{gravwils}
\end{equation}
In a purely classical context, there is no specific need to consider the trace, since the loop holonomy can be thought
of as an invariant (1,1)-tensor. Taking the trace is primarily motivated by quantum considerations, where neither the location
of a specific base point nor the choice of a specific frame are meaningful concepts in an ensemble average 
over spacetime geometries. 
As already mentioned in the introduction, there are various ways in which one can envisage turning the gravitational Wilson loop into
a genuine quantum observable, for example, by performing some averaging over its location in space, or by marking its location
with external matter. The measurement of a Wilson loop observable in four-dimensional CDT quantum gravity in \cite{wilsoncdt} 
considered an ensemble of loops along the worldline of a massive particle, identified cyclically in time.

Since the holonomies of loops on general $d$-dimensional Riemannian manifolds take values in $SO(d)$, the corresponding Wilson loops
can be parametrized by $r$ angles $\alpha_j$, $j=1,\dots,r$, where $r$ equals the rank of the group, 
given by $d/2$ for even $d$ and $(d-1)/2$ for odd $d$. 
In three dimensions, the Wilson loop has the form
\begin{equation}
\mathrm{Tr}\,W_\gamma=1+2\cos(\alpha_\gamma),
\label{3dWilson}
\end{equation}
where $\alpha_\gamma$ is the angle of rotation around some fixed axis, and it should be kept in mind 
that this quantity has a nonlocal
functional dependence on the loop $\gamma$ and the field $A$ along it.
In four dimensions, the Wilson loop depends on two angles, 
\begin{equation}
\mathrm{Tr}\,W_\gamma=2\cos(\alpha_\gamma)+2\cos(\beta_\gamma),
\label{4dWilson}
\end{equation}
parametrising independent rotations in two mutually orthogonal planes.

Our central aim will be to express the gauge-invariant content of the gravitational holonomy and 
the Wilson loop \rf{gravwils} in terms of surface integrals over local curvature and to identify conditions 
under which this can be done. 
To express the angles $\alpha_\gamma$ and $\beta_\gamma$ as functions of curvature we first need to understand 
the behaviour of the corresponding holonomies.
As mentioned earlier, a key obstacle to computing holonomies like eq.\ \rf{wline}
on a general curved manifold $(\mathcal{M}, g_{\mu\nu})$ in dimension $d\geq 3$ and relating them to surface integrals
is the nonabelian character of the field $A(t)$. We will treat the cases $d\! =\! 3$ and $d\! =\! 4$ in Sec.\ \ref{tgsurfaces} below, 
without requiring that 
the holonomy group of the manifold $\cal M$ be abelian, which would render the discussion trivial. 
Before turning to this explicit construction, we will in the next section review the derivation of the nonabelian Stokes' theorem 
for Riemannian manifolds, since it contains some elements we will be needing later.

\section{Geometric construction of the nonabelian Stokes' theorem}
\label{nasty}

As outlined above, our main aim is to retrieve information about the curvature 
of a manifold from measuring finite holonomies or their associated Wilson loops. This is motivated 
by the well-known relation between local curvature at a point $p\in {\cal M}$ and the holonomy of an {\it infinitesimal} loop based at $p$.
Let us recap briefly how this comes about. 
Consider two linearly independent tangent vectors $v$ and $w$ in $T_p{\cal M}$, which for simplicity are chosen mutually
orthogonal. They determine a two-dimensional surface $S$ locally, consisting of geodesics starting at $p$ whose tangent vector at $p$ lies in
the span of $v$ and $w$. Next, set up a local Riemann normal coordinate system $\{x_i\}$ based at $p=(0,0,\dots)$, 
such that $v$ and $w$ point along the positive $x_1$- and $x_2$-direction, respectively. 
For sufficiently small $\varepsilon >0$, one can construct geometrically a small square surface of linear size $\varepsilon$ in $S$,
which consists of all points $x^\mu(\sigma,\tau)=(\sigma,\tau,\vec 0)$, $\sigma,\tau\in [0,\varepsilon]$, 
with $\vec 0$ denoting the $d-2$ vanishing coordinates in the remaining directions.

\begin{figure}
\centering
\includegraphics[width=0.4\textwidth]{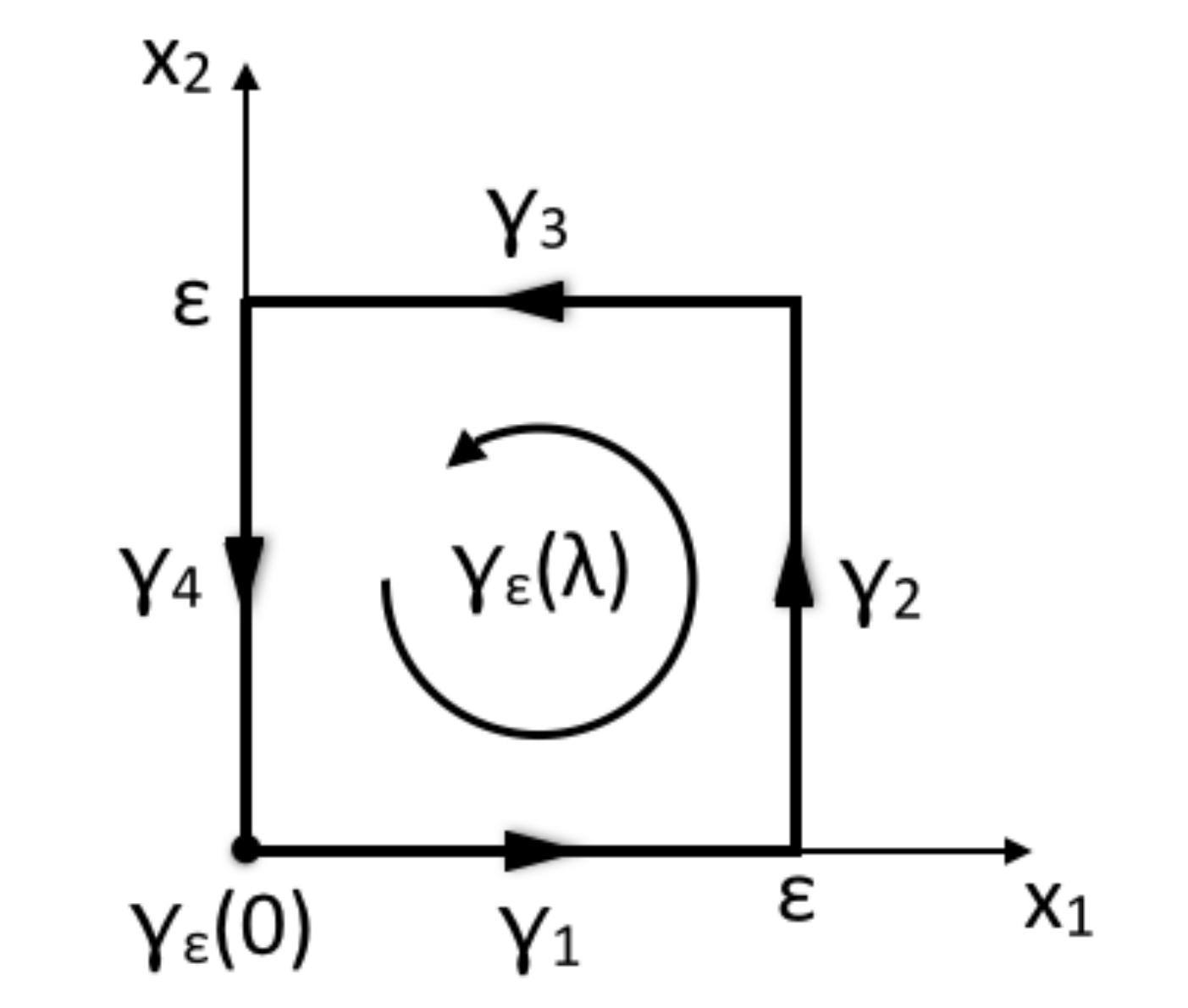}
\caption{\label{fig:plaquette} Infinitesimal square loop $\gamma_\varepsilon (\lambda)$ of side length $\varepsilon$ 
in Riemann normal coordinates $x_1$ and $x_2$, as described by eqs.\ (\ref{infsq}).}
\end{figure}

One can use the same coordinates to describe a closed oriented path $\gamma_\varepsilon (\lambda)$ 
that runs along the boundary of the small square, 
starting from the origin in positive $x_1$-direction (see Fig.\ \ref{fig:plaquette}).
The loop $\gamma_\varepsilon(\lambda)$ can be decomposed into path segments along the four sides of the square,
$\gamma_\varepsilon =\gamma_4\circ\gamma_3\circ\gamma_2\circ\gamma_1$, which in terms of the Riemann normal coordinates
of the local 1-2-``plane" can be parametrized by
\begin{eqnarray}
&&\gamma_1(\lambda)=(\lambda,0),\;\;\;\;\;\;\;\;\; \lambda\in[0,\varepsilon],\;\; \nonumber\\ 
&&\gamma_2(\lambda)=(\varepsilon, \lambda-\varepsilon),\;\;\; \lambda\in [\varepsilon,2\varepsilon],\;\; \nonumber\\
&&\gamma_3(\lambda)=(3\varepsilon-\lambda,\varepsilon),\; \lambda\in [2\varepsilon, 3\varepsilon],\;\; \nonumber\\
&&\gamma_4(\lambda)=(0,4\varepsilon - \lambda),\; \lambda\in  [3\varepsilon, 4\varepsilon].
\label{infsq}
\end{eqnarray}

Following eq.\ (\ref{loopcomp}), the holonomy $W_{\gamma_\varepsilon}$ around the infinitesimal square is given by the product of the
individual edge holonomies,
\begin{equation}
W_{\gamma_\varepsilon}=U_{\gamma_4}U_{\gamma_3}U_{\gamma_2}U_{\gamma_1}.
\label{loopsqu}
\end{equation}
To exhibit the dependence of this holonomy on the Riemann tensor, one can now perform an $\varepsilon$-expansion
of (\ref{loopsqu}), for example, by using the nested integral form (\ref{holoexpand}) of the holonomy as a starting point, 
substituting the field $A$ by the contraction of the Levi-Civita connection with the tangent vector to $\gamma_\varepsilon$. 
To lowest nontrivial order in $\varepsilon$, one finds
\begin{equation}
(W_{\gamma_\varepsilon})^\kappa{}_\lambda=\mathbb{1}^\kappa{}_\lambda- 
\varepsilon^2 R^\kappa{}_{\lambda 12}(p)+o(\varepsilon^2)\equiv ({\rm e}^{-\varepsilon^2 R_{12}})^\kappa{}_\lambda
+o(\varepsilon^2).
\label{squareexp}
\end{equation}  
This illustrates that all information about the curvature at $p$ can be obtained by considering holonomies associated
with infinitesimal square loops in general $i$-$j$-planes through $p$.

There have been attempts to generalize the result (\ref{squareexp}) to a noninfinitesimal surface $S$,
by relating the loop holonomy $W_\gamma$ of the loop $\gamma$ with image $\partial S$
to a curvature integral over $S$. In the context of Yang-Mills theory, this usually goes by the name of 
{\it nonabelian Stokes' theorem}. An example is the classic reference \cite{aref}, where one re-expresses
the (path-ordered) holonomy of a nonabelian $su(N)$-gauge connection $A$ of a square loop on flat, Minkowskian spacetime 
in terms of a two-dimensional {\it surface-ordered} integral of the corresponding nonabelian field strength tensor $F[A]$ over the surface 
enclosed by $\gamma$ (for related later references, see e.g. \cite{fishbane,gross}).

However, there is a much earlier and apparently little-known discussion of the Riemannian case and the 
Levi-Civita connection due to Schlesinger \cite{schlesinger}, which is more relevant to our case and 
closely parallels the more recent physics applications involving gauge connections in QCD.  
The key result one is after in both cases is an analogue of Stokes' theorem 
for surfaces, 
\begin{equation}
\oint_{\partial S} A = \int_S dA ,
\label{stokesnorm}
\end{equation}
for the case that the smooth one-form $A$ and the curvature two-form, given by its exterior derivative $dA$, are matrix-valued.
Recall that Stokes' theorem also holds for surfaces with corners and that the orientation of $\partial S$ must be the appropriate 
one induced from the orientation of $S$ such that eq.\ (\ref{stokesnorm}) is valid without a relative minus sign between
the two sides of the equation (see, for example, \cite{leeintro}).%
\begin{figure}[t]
\centering
\includegraphics[width=0.85\textwidth]{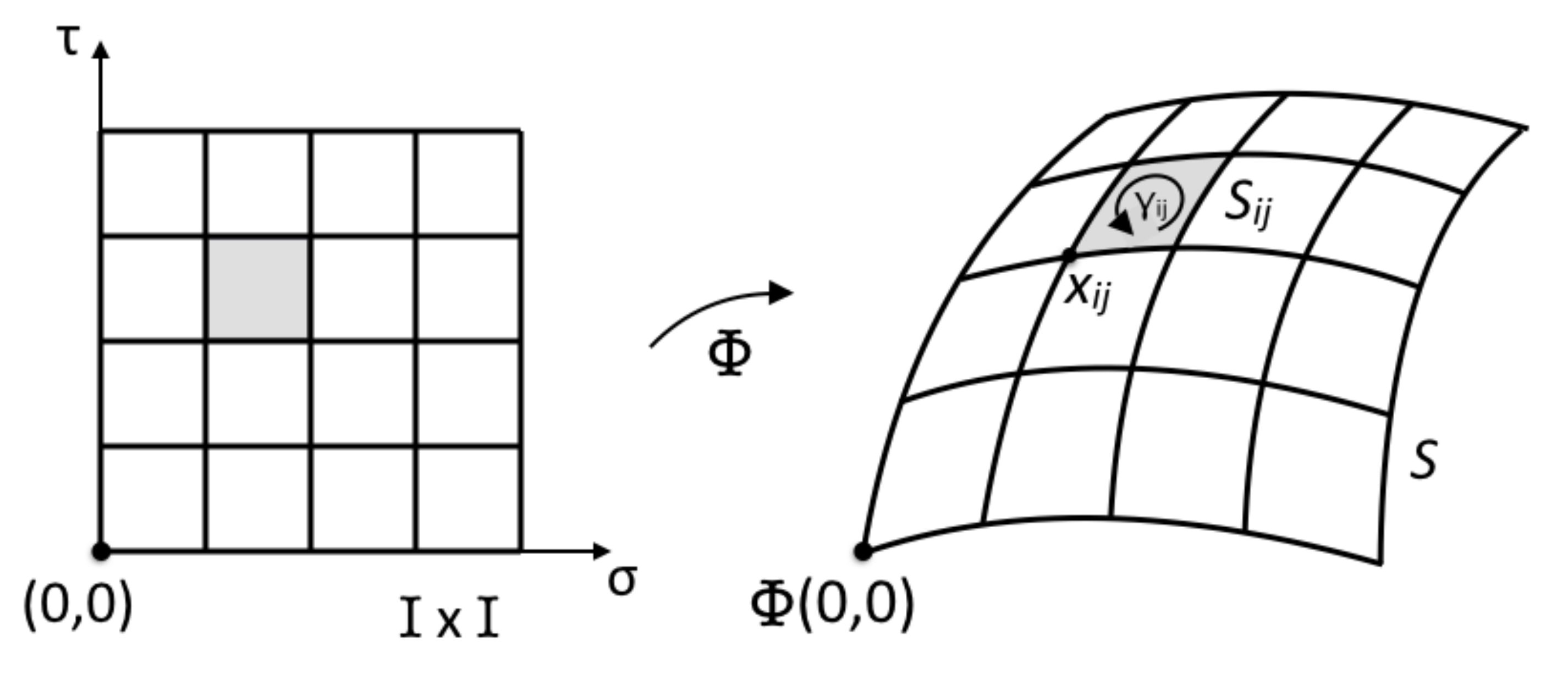}
\caption{\label{fig:grid} The surface $S$ embedded in $\cal M$ inherits a grid structure from the flat unit square $I\times I$,
consisting of $N^2$ plaquettes $S_{ij}$, $i,j=0,\dots N-1$. Each of them is associated with a plaquette loop $\gamma_{ij}$, based at
$x_{ij}$ and running counterclockwise along its boundary $\partial S_{ij}$.}
\end{figure}
We will review briefly the derivation of the generalized Stokes' theorem for Riemannian manifolds, because we will use
elements of the proof in our construction in Sec.\ \ref{tgsurfaces} of an alternative relation between Wilson loops and curvature. 
Besides the original paper \cite{schlesinger}, a useful reference for this material is \cite{chafom}, 
which deals with the closely related case where the one-form $A$ 
takes values in a finite-dimensional matrix Lie algebra. The nonabelian character of this algebra makes the construction
nontrivial. Analogous to the need for path ordering when integrating a nonabelian connection along a one-dimensional
path, as in eq.\ (\ref{wline}) above, integrating a nonabelian curvature over a two-dimensional surface will require a prescription of surface ordering. In addition, one needs a scheme of parallel transport to a common base point at which the 
contributions from individual infinitesimal surface elements can be composed by matrix multiplication. Unfortunately, the construction
of the nonabelian Stokes' theorem does not suggest a useful notion of averaged or coarse-grained curvature, at least none
we have been able to discern. This has motivated us to look for an alternative prescription, where under suitable circumstances
holonomies or Wilson loops capture average curvature in a more straightforward way.

In order to establish the nonabelian Stokes' theorem, suppose we are given a smooth embedding 
$\Phi(\sigma,\tau):U\rightarrow {\cal M}$ from some open subset $U\subset\mathbb{R}^2$ into the manifold $\cal M$. 
Consider a unit square $I\times I\subset U$ parametrized by $\sigma,\,\tau \in [0,1]$ and contained in $U$. 
Its image in $\cal M$ under the embedding is a ``rectangular" surface\footnote{We mean a surface with four corners, 
not a metric rectangle in flat Euclidean space.} $S$, whose boundary $\partial S$ is the
image of the boundary of $I\times I$. We will consider a tiling of the surface by $N\times N$ elementary 
rectangles $S_{ij}$ -- referred to as ``plaquettes" in what follows -- 
whose lower left-hand corner is located at the point $x_{ij}\! :=\!\Phi (i/N,j/N)$, $i,j=0,\dots,N-1$ (see Fig.\ \ref{fig:grid}).
The small rectangular loop 
based at $x_{ij}$ and running counterclockwise along the boundary $\partial S_{ij}$ of the plaquette $S_{ij}$
will be called a ``plaquette loop" and denoted by $\gamma_{ij}$. It is the image 
in $\cal M$ under the map $\Phi$ of a small square loop in $I\times I$ with edge length $1/N$, based at 
$(\sigma,\tau)\! =\! (i/N,j/N)$, whose straight edges lie along coordinates lines of constant $\tau$ or constant $\sigma$. 
Its associated holonomy $W_{ij}$ is related to the Riemann curvature tensor $R$ at $x_{ij}$ by
\begin{equation}
W_{ij}:=W_{\gamma_{ij}}=\mathbb{1}-\frac{1}{N^2} R(\dot{\Phi},\Phi')+o(N^{-2}) =\mathbb{1}-\frac{1}{N^2} 
R(x_{ij})^.{}_{. \mu\nu}\dot{\Phi}^\mu(x_{ij})\Phi'^\nu(x_{ij})+o(N^{-2}),
\label{plaqhol}
\end{equation} 
where the tangent vectors in the two coordinate directions are defined as 
$\dot{\Phi}\! :=\!\partial\Phi/\partial \sigma$ and $\Phi'\! :=\!\partial\Phi/\partial\tau$.

\begin{figure}
\centering
\includegraphics[width=0.9\textwidth]{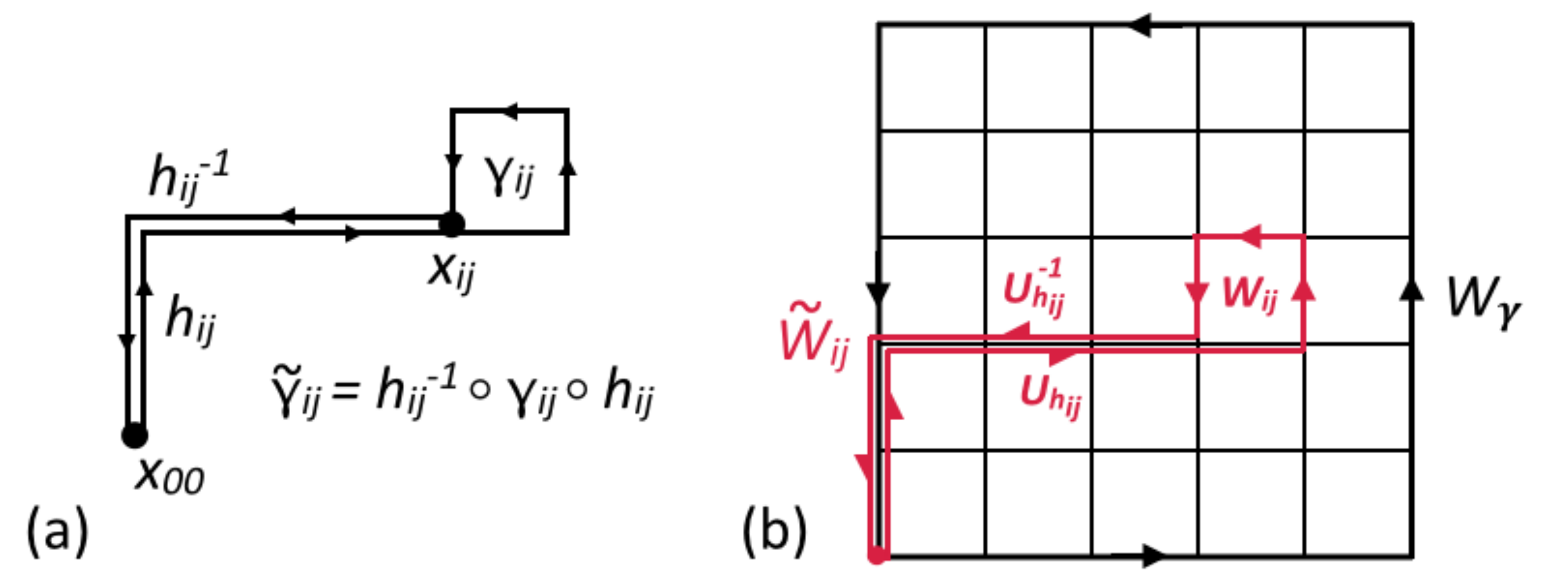}
\caption{\label{fig:plaqholo} (a): To compose the plaquette holonomies $W_{ij}$ at the common base point
$x_{00}$, we must choose for each plaquette loop $\gamma_{ij}$ a path $h_{ij}$ connecting $x_{00}$ and $x_{ij}$.
(b): The holonomy $\tilde{W}_{ij}$ of the resulting loop $\tilde{\gamma}_{ij}$ is obtained by multiplying together the
individual holonomies $W_{ij}$, $U_{h_{ij}}$ and $U_{h_{ij}}^{-1}$.}
\end{figure}
Loop holonomies associated with different plaquette loops cannot be composed because they are based at different points
and therefore refer to different tangent spaces.
However, we can parallel-transport them to a common base point, which we choose to be $x_{00}=\Phi (0,0)$, and compose 
them there. This involves the choice of a path $h_{ij}$ in $S$ connecting the base point $x_{00}$ to the origin $x_{ij}$ of a given
plaquette loop $\gamma_{ij}$, leading to a closed path
\begin{equation}
\tilde{\gamma}_{ij}=h_{ij}^{-1}\circ\gamma_{ij}\circ h_{ij}
\label{gammatilde}
\end{equation}
based at $x_{00}$. The associated loop holonomy will be denoted by $\tilde{W}_{ij}$, where
\begin{equation}
\tilde{W}_{ij}:=W_{\tilde{\gamma}_{ij}}=U_{h_{ij}}^{-1}\, W_{ij}\, U_{h_{ij}}.
\label{wtilde}
\end{equation}
Our choice for the piecewise smooth path $h_{ij}$ in $\cal M$ is the image under the map $\Phi$ of the path in
$I\times I$ that starts at $(\sigma,\tau)=(0,0)$, proceeds along the $\tau$-axis to the point $(0,j/N)$ and from
there runs parallel to the $\sigma$-axis until it reaches the point $(i/N,j/N)$, see Fig.\ \ref{fig:plaqholo}. 
Expanding the parallel-transported plaquette holonomy for small $1/N$ gives 
\begin{equation}
\tilde{W}_{ij}=\mathbb{1}-\frac{1}{N^2}\, U_{h_{ij}}^{-1}\, R(\dot{\Phi},\Phi')\, U_{h_{ij}}+o(N^{-2})=:
\mathbb{1}-\frac{1}{N^2}\, \tilde{R}(x_{ij})+o(N^{-2}),
\label{plaqtilde}
\end{equation}
from conjugating eq.\ (\ref{plaqhol}) by the path holonomy $U_{h_{ij}}$. Note that the parallel-transported cur\-vature
$\tilde{R}(x_{ij})\! :=\! U_{h_{ij}}^{-1}\, R(x_{ij})\, U_{h_{ij}}$ in eq.\ (\ref{plaqtilde}) is no longer a local expression 
at the point $x_{ij}$, but also depends in a nonlocal way on the path $h_{ij}$. (For compactness we
suppress this dependence in the notation.)
Moreover, $\tilde{R}(x_{ij})$ acts by $SO(d)$-rotation on tangent vectors at the base point $x_{00}$.

The next step in the construction is to compose the $N^2$ plaquette holonomies $\tilde{W}_{ij}$ in such a way 
that their ordered product is equal to $W_\gamma$, where $\gamma$ denotes the closed path running along the boundary
$\partial S$ of the original surface $S$ in counterclockwise direction. To demonstrate that this is possible recall 
from our remarks after eq.\ (\ref{ptrans})
that if in the composition of loops $\tilde{\gamma}_{mn}\circ\tilde{\gamma}_{kl}\circ\dots\circ \tilde{\gamma}_{ij}$
some path segment $p$ is followed immediately by the path segment $p^{-1}$ with opposite orientation, their associated holonomies 
are inverses of each other and therefore cancel, $U_{p^{-1}} U_p=\mathbb{1}$.
There are many ways of choosing an ordering prescription for the holonomies $\tilde{W}_{ij}$ such that all holonomies 
associated with internal path segments in $S$ cancel. Any particular choice determines the type of
surface ordering that will appear on the right-hand side of the nonabelian Stokes' theorem.           

\begin{figure}
\centering
\includegraphics[width=0.4\textwidth]{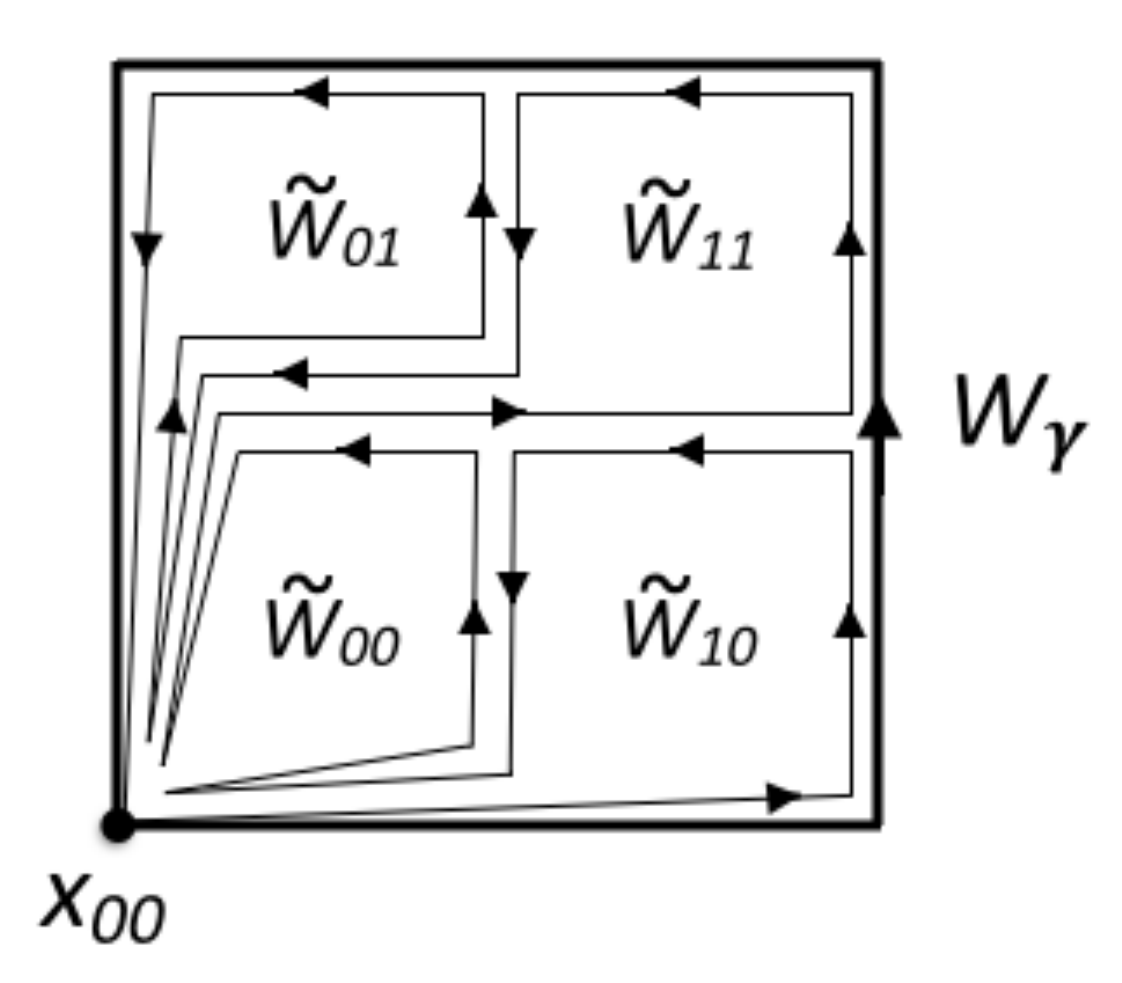}
\caption{\label{fig:plaqcombi} Simple example of the ordering prescription for the parallel-transported plaquette holonomies
$\tilde{W}_{ij}$, for the case $N=2$. The total holonomy is obtained by composing the four plaquette holonomies in the order
indicated, $W_\gamma = \tilde{W}_{01}\tilde{W}_{11}\tilde{W}_{00}\tilde{W}_{10}$. Holonomies associated with 
``backtracking" internal lines cancel each other.}
\end{figure}
The simple ordering prescription we will adopt starts from the plaquette with labels $(i,j)=(N-1,0)$ in the lower right-hand corner, 
and then moves to the left until the plaquette $(0,0)$ is reached and the bottom row is filled. One then proceeds
to the plaquette on the far right in the second row from the bottom, labelled $(N-1,1)$, followed by the plaquette $(N-2,1)$
to its left, and continues until this row is filled (see Fig.\ \ref{fig:plaqcombi} for illustration). 
This process is repeated by filling each subsequent row from right
to left until finally the last row is completed. It is convenient to introduce the ``row holonomy" $\tilde{W}_{{\cal R}_j}$ for the $j$th
row, $j=0,\dots,N-1$, by multiplying together the individual plaquette holonomies in the way just described,
\begin{equation}
\tilde{W}_{{\cal R}_j}:= \tilde{W}_{0j}\, \tilde{W}_{1j}\, \dots \tilde{W}_{N-2,j}\, \tilde{W}_{N-1,j}.
\label{rowholo}
\end{equation}
One easily convinces oneself that with the given ordering the holonomies of all path segments internal to
the outer loop $\gamma$ cancel, leading to the desired identity
\begin{equation}
W_\gamma = \tilde{W}_{{\cal R}_{N-1}}\, \tilde{W}_{{\cal R}_{N-2}}\, \dots \tilde{W}_{{\cal R}_1}\, \tilde{W}_{{\cal R}_0}.
\label{multirow}
\end{equation}
The challenge now is to demonstrate that relation (\ref{multirow}) can be rewritten such that it assumes
the form of (a nonabelian, exponentiated version of) Stokes' theorem, eq.\ (\ref{stokesnorm}). The left-hand side
$W_\gamma= {\rm \mathrm{P} \,  e}^{-\oint_\gamma \Gamma}$ already has the desired form, but for the right-hand side
this remains to be shown. 

By a judicious rearrangement of the terms appearing in the joint expansions of the type (\ref{holoexpand}) and
(\ref{holoalt}) of the plaquette holonomies contributing to a given row \cite{chafom,aref}, and taking a continuum
limit in the $\sigma$-direction (allowed by virtue of the absolute convergence of all expressions), the row holonomy 
$\tilde{W}_{{\cal R}_j} $ can be rewritten as 
\begin{equation}
\tilde{W}_{{\cal R}_j} = \mathbb{1}-\frac{1}{N} \int_0^1\!\! d\sigma\, \tilde{R}(\Phi(\sigma,j/N))+o(N^{-1})=: 
\mathbb{1}-\frac{1}{N}\, {\mathcal A}(j/N)+o(N^{-1}).
\label{rowlimit}
\end{equation}
It turns out that only terms up to order $1/N$ will contribute to the final result. Note that there are no factor-ordering 
ambiguities at this order, but that row holonomies (\ref{rowlimit}) for different values of $j$ will in general {\it not} commute
with each other. The new symbol $\cal A$ has been adopted for compactness of notation only.
Next, we apply the standard path ordering -- in the sense of the earlier introduced product integration -- to the product
of the noncommuting row holonomies in the $\tau$-direction, to obtain
\begin{equation}
\lim_{N\rightarrow\infty} (\mathbb{1}- \frac{1}{N}\, {\mathcal A}\left( \frac{\scriptstyle N-1}{\scriptstyle N} \right) )
(\mathbb{1}- \frac{1}{N}\, {\mathcal A}\left( \frac{\scriptstyle N-2}{\scriptstyle N} \right) )
\dots
(\mathbb{1}- \frac{1}{N}\, {\mathcal A}\left( \frac{\scriptstyle 1}{\scriptstyle N} \right) )
(\mathbb{1}-\frac{1}{N}\, {\mathcal A}(0))= {\rm \mathrm{P}}\exp ({-\!\!\! \int\!\! d\tau\, {\mathcal A}(\tau) })
\label{holotot}
\end{equation}
in the limit of infinite $N$, where the integrand ${\mathcal A}(\tau)$ in the exponent is itself of the form of an integral,
\begin{equation}
{\mathcal A}(\tau)=\int_0^1 \!\! d\sigma\, \tilde{R}(\Phi(\sigma,\tau)).
\label{Aint}
\end{equation}
Introducing a new symbol ``$\mathcal P$" to indicate surface or area ordering, the right-hand side
of eq.\ (\ref{holotot}) can be written as 
\begin{equation}
{\mathcal P}\exp \left( -\!\! \int_0^1\!\! d\tau\int_0^1 \!\! d\sigma\, \tilde{R}(\Phi(\sigma,\tau)) \right).
\label{areaord}
\end{equation}
The definition of $\mathcal P$ is inherited directly from our choice of plaquette ordering: 
for factors of $\tilde{R}$ with identical $\tau$-argument, the one with the smaller $\sigma$-argument always
appears ordered to the left, while for factors of $\tilde{R}$ with different $\tau$-argument, the one with the bigger
$\tau$-argument always appears ordered to the left, independent of the values of $\sigma$. 
However, as already noted in connection with eq.\ (\ref{rowlimit}), it turns out
that with our particular choice of plaquette ordering the ordering ambiguities with respect to the parameter $\sigma$ 
drop out in the limit as $N \rightarrow\infty$.

Writing the expression (\ref{areaord}) as an infinite expansion analogous to eq.\ (\ref{holoexpand}), 
\begin{equation}
(\ref{areaord})= \mathbb{1} - \!\! \int_0^1\!\! d\tau\!\int_0^1 \!\! d\sigma\, \tilde{R}(\Phi(\sigma,\tau)) +
\int_0^1\!\! d\tau_2\! \int_0^{\tau_2}\!\!\!\! d\tau_1\! \int_0^1 \!\! d\sigma_2 \! \int_0^1 \!\! d\sigma_1 \,
\tilde{R}(\Phi(\sigma_2,\tau_2))  \tilde{R}(\Phi(\sigma_1,\tau_1)) +\dots, 
\label{areaexpand}
\end{equation}
the surface ordering is reflected in the nontrivial integration limit of the parameter $\tau_1$ in the third term, with
a similar nesting of the multiple integrals in the higher-order terms not written explicitly in eq.\ (\ref{areaexpand}).

Putting together eqs.\ (\ref{multirow}), (\ref{rowlimit}), (\ref{holotot}) and (\ref{areaord}), we have finally arrived 
at the nonabelian Stokes' theorem, which schematically reads 
\begin{equation}
{\rm \mathrm{P} \,  e}^{-\oint_{\partial S} \Gamma}={\rm \mathcal{P} \,  e}^{-\int_S \tilde{R}}.
\label{nast}
\end{equation}
Let us recapitulate how this relation differs from its standard, abelian counterpart. Firstly, (\ref{nast}) is a matrix-valued
equation, reflecting the nontrivial index structure of the Levi-Civita connection $\Gamma$ and its associated curvature $R$. 
Secondly, the nonabelian nature of the holonomy group requires a path ordering on the left-hand side and a surface ordering on the
right-hand side of the identity. While the former is unique for a given path, the surface ordering is not. However, 
the invariance of (\ref{nast}) under a change of surface ordering scheme is not manifest. For example, even the
rather simple scheme we used treats the $\sigma$- and $\tau$-direction in a very asymmetric manner. Thirdly,
the construction leading to (\ref{nast}) depends on a base point $\Phi(0,0)$, and both sides still transform nontrivially under
a change of frame at this point. If desired, the dependence on this base point can of course be removed by taking the matrix trace
on both sides of the equation. Fourthly, although the left-hand side of (\ref{nast}) only depends on the boundary $\partial S$, 
the invariance of the right-hand side
under smooth deformations of $S$ in $\mathcal M$ leaving $\partial S$ invariant (for the case $d\geq 3$) is not obvious.   
Lastly, and most relevant for our purposes, the expression in the exponent of the right-hand side of the nonabelian
Stokes' theorem is {\it not} an ordinary surface integral of a local integrand. Instead, $\tilde R$ encodes information about the
Riemann tensor in a complicated, nonlocal and seemingly path-dependent way. In general, this makes it difficult to interpret the 
right-hand side of (\ref{nast}) in terms of an averaged or coarse-grained curvature or some other familiar quantity.  
Unsurprisingly, the situation becomes much simpler when the holonomy group is abelian, as happens for example in $d=2$. In this
case there is no need for path or surface ordering, and the integrand of the surface integral becomes local. 

This raises the question of whether and under what circumstances one may be able to establish a relation between 
the holonomy of a finite loop and some notion of integrated or averaged curvature on a manifold with $d\geq 3$ 
whose holonomy group is not necessarily abelian.
In the next section, we will construct such a relation, starting in $d=3$. It is still not applicable to general Riemannian manifolds, but
only those that possess special symmetries, and only holds for a selected class of loops. However, unlike the nonabelian Stokes'
theorem, its interpretation in terms of an integrated local curvature and its invariance under smooth surface deformations will be
explicit and straightforward.

\section{Holonomies on totally geodesic surfaces}
\label{tgsurfaces}

We saw in the previous section that the nonabelian character of the (restricted) holonomy group $SO(d)$ for a general Riemannian
manifold with $d>2$ leads to serious complications when trying to interpret the holonomy of a non-infinitesimal loop as a measure of
average curvature. As already alluded to in the introduction, this can be seen as part of a more general problem, that of averaging 
tensorial quantities on a curved space where parallel transport is nontrivial. In the derivation of the nonabelian Stokes' theorem,
this was reflected in the need for choosing a particular, nonunique surface ordering, obscuring the geometric meaning of the result. 
While it is interesting and nontrivial that this theorem can be derived, it does not seem to be of practical use for
our purposes, and to the best of our knowledge has only found limited application elsewhere in physics or mathematics.   

In the following, we will focus on a particular class of curved Riemannian manifolds, which contain so-called totally geodesic surfaces.
There are many examples of such manifolds, but they are not generic. Loosely speaking, they possess special symmetry properties, 
examples of which will be given below. Their holonomy groups are not necessarily abelian, but the crucial property we will use is
that if a loop is confined to lie completely inside a totally geodesic surface, its holonomy assumes a particularly simple form. 
For the cases that have our particular interest, namely, $d=3$ and $d=4$, the holonomy turns out to be abelian,   
and we can relate it to curvature-dependent surface expressions that do not need any surface ordering. -- After introducing
totally geodesic surfaces and their geometric properties, we will examine the three-dimensional case in Sec.\ \ref{sec:d3}. 
We relate the holonomy of a loop $\gamma$ in a totally geodesic surface $\mathcal{S}_{tg}$ with a curvature integral over the 
part $S\subset \mathcal{S}_{tg}$ of the surface that is enclosed by $\gamma\sim \partial S$. 
In Sec.\ \ref{sec:threedim} we demonstrate that this curvature 
integral is invariant under smooth deformations of $S$ away from the totally geodesic surface $\mathcal{S}_{tg}$, but which
leave $\gamma$ fixed. Sec.\ \ref{sec:S3} illustrates the construction with a nontrivial example.

A two-dimensional manifold $\mathcal S$ embedded in a $d$-dimensional Riemannian manifold $(\mathcal{M},g_{\mu\nu})$,
together with the metric $\tilde{g}_{\mu\nu}$ induced by $g_{\mu\nu}$, constitutes a Riemannian submanifold of $\mathcal{M}$. Such a
submanifold is called a {\it totally geodesic surface} if every geodesic of $(\mathcal{S},\tilde{g}_{\mu\nu})$ is also a geodesic of 
the ambient manifold $(\mathcal{M},g_{\mu\nu})$.\footnote{Totally geodesic submanifolds are covered by many textbooks
and standard references on Riemannian geometry, see, for example, \cite{heckman}.} It follows that parallel
transport in $\mathcal M$, when applied to curves $\gamma$ in $\mathcal{S}_{tg}$, preserves the splitting 
\begin{equation}
T\mathcal{M} =T\mathcal{S}_{tg}\oplus N\mathcal{S}_{tg}
\label{split}
\end{equation}
of the tangent space to $\mathcal{M}$ into the direct sum of the tangent space to $\mathcal{S}_{tg}$ and its orthogonal complement 
$N\mathcal{S}_{tg}$. In other words, parallel transport of a vector $v\in T_p\mathcal{S}_{tg}$ along a closed curve $\gamma$ in 
$\mathcal{S}_{tg}$ will result in another tangent vector $v'\in T_p\mathcal{S}_{tg}$, while parallel transport of any vector 
$n\in N_p\mathcal{S}_{tg}$ along the same curve will result in another normal vector $n'\in N_p\mathcal{S}_{tg}$ \cite{cartan,berger}. 
Accordingly, the holonomy $W_\gamma$ of a loop $\gamma$ in $\mathcal{S}_{tg}$ takes values in the direct product, 
\begin{equation}
W_\gamma \in SO(2) \times SO(d-2) \subset SO(d),
\label{SubGroup}
\end{equation}
which is an {\it abelian} group for both $d=3$ and $d=4$. 

Totally geodesic surfaces or, more generally, totally geodesic submanifolds occur in Riemannian manifolds with 
symmetries \cite{berger}.
For example, spaces of constant curvature can have many totally geodesic submanifolds.
Also, any connected component of the fixed point set of an involutive isometry $s$ of $(\mathcal{M},g_{\mu\nu})$ 
(an isometry whose square is the identity, 
$s\circ s= Id$) is a totally geodesic submanifold. If the manifold $\mathcal M$ is flat Euclidean space, all two-planes are
totally geodesic surfaces, as follows immediately from both the criterion for geodesics and the one on involutive isometry.
For general, curved manifolds $\mathcal M$, totally geodesic surfaces are the closest analogues of such planes in terms of
their curvature properties as embedded subspaces. These properties are conveniently captured by the so-called
second fundamental form $II$ associated with the surface.  

For a given two-dimensional surface $\mathcal{S}$, the second fundamental form provides a measure for the
difference between geodesics in $\mathcal{S}$ with respect to the induced metric $\tilde g$ and geodesics in the ambient
space $(\mathcal{M},g)$ that share the same initial conditions. 
At a given point $p\in\mathcal{S}$ the second fundamental form is a 
symmetric bilinear form that maps a pair of tangent vectors into a normal vector according to 
\begin{equation}
II_p: T_p \mathcal{S}\times T_p \mathcal{S}\rightarrow N_p\mathcal{S}, \;\; II(v,w):= \nabla_v w-\tilde{\nabla}_v w,
\label{sff}
\end{equation}
where $\nabla$ and $\tilde\nabla$ denote the covariant derivatives associated with the metrics $g$ and $\tilde g$ respectively. 
Choosing a local orthonormal basis of vectors $\hat{n}_i$, $i \in \lbrace 1, ..., d-2\rbrace$
for the $(d-2)$-dimensional normal space $N\mathcal{S}$ in a neighbourhood of $p$ in $\mathcal S$, we can write
the second fundamental form as
\begin{equation}
II(v,w)= \sum^{d-2}_{i=1} \langle \hat{n}_i, \nabla_v w\rangle \hat{n}_i, \ \ \ \ v, w \in T \mathcal{S},
\label{sform}
\end{equation}
where $\langle\cdot,\cdot\rangle=g(\cdot,\cdot)$ denotes the inner product on $T\mathcal{M}$.
From the definition of a totally geodesic surface in terms of the behaviour of its geodesics 
it follows that a surface $\mathcal S$ is totally geodesic if and only if its second 
fundamental form vanishes identically, 
\begin{equation}
II_p(v,w)=0,\;\; \forall v,w\in T_p \mathcal{S},\; \forall p\in\mathcal{S}.
\label{sffnull}
\end{equation}
Note that being a totally geodesic surface is a stronger condition than being a minimal surface, which would require only the
trace of the second fundamental form to be zero. 
The vanishing of $II$ implies that the Riemann curvature tensor $\tilde{R}$ of a totally geodesic surface $\mathcal{S}_{tg}$ 
coincides with the Riemann curvature tensor $R$ of the ambient space when evaluated on vectors tangent to $\mathcal{S}_{tg}$,
\begin{equation}
\langle R(u,v)w,z\rangle = \langle \tilde{R}(u,v)w,z\rangle,\;\;\; u,v,w,z \in T\mathcal{S}_{tg}.
\label{riem2}
\end{equation}

In the following subsection we will use the abelian nature of the holonomies on totally geodesic surfaces in three-dimensional
Riemannian manifolds $\mathcal{M}$ to relate the holonomy to a surface integral of curvature, in the spirit of the
construction of the nonabelian Stokes' theorem, but without the complications arising from noncommutativity.

\subsection{Holonomies on totally geodesic surfaces in three dimensions}
\label{sec:d3}

The holonomies of contractible loops in three dimensions take values in the rotation group $SO(3)$,
whose elements are characterised by a rotation axis in $\mathbb{R}^3$ and a rotation angle around that axis. 
We will now consider holonomies of curves $\gamma$ with base point $p$ that lie inside a totally geodesic 
hypersurface $\mathcal{S}_{tg}$ of some three-dimensional Riemannian manifold $\mathcal M$. 
We can always choose local coordinates $(x_1,x_2,x_3)$ in a neighbourhood of $p$ such that 
$\mathcal{S}_{tg}$ is given by $x_3=0$ and the normal vector $\hat n$ to the surface points along the positive
$x_3$-direction. It follows from our discussion in Sec.\ \ref{tgsurfaces} above that the holonomy matrix of any closed loop
$\gamma(\tau)$ in $\mathcal{S}_{tg}$ based at the point $p\in \mathcal{S}_{tg}$ has the form
\begin{equation}
W_\gamma =\left(\begin{array}{ccc}\cos\alpha_\gamma & -\sin\alpha_\gamma & 0 \\
\sin\alpha_\gamma & \cos\alpha_\gamma & 0 \\0 & 0 & 1\end{array}\right)
\label{rotmat}
\end{equation}
with respect to an orthonormal basis of the tangent space $T_p\mathcal{M}$, where the angle $\alpha_\gamma$ is
measured counterclockwise from the $x_1$-axis in the $x_1$-$x_2$-plane. In other words, all such holonomies lie in
the $SO(2)$-subgroup of rotations that leave the normal direction to the surface $\mathcal{S}_{tg}$ invariant.
As a result, the holonomies of any pair $\gamma_1$, $\gamma_2$ of such loops commute, 
\begin{equation}
[W_{\gamma_1},W_{\gamma_2}]=0,\;\;\;\; \gamma_1(\tau), \gamma_2(\tau) \subset \mathcal{S}_{tg}.
\label{commuting}
\end{equation}
While the form (\ref{rotmat}) of the holonomy matrix still depends on the choice of basis in $T_p\mathcal{M}$,
its trace
\begin{equation}
\mathrm{Tr}\, W_\gamma=1+2\cos \alpha_\gamma
\label{trwg}
\end{equation}
does not. It allows us to extract the rotation angle $\alpha_\gamma$ up to a sign.  
    
Next, we follow the steps of the derivation of the nonabelian Stokes' theorem in Sec.\ \ref{nasty} to
establish a relation between the holonomy of a finite-sized rectangular loop $\gamma\subset \mathcal{S}_{tg}$
and a curvature integral over the surface $S_{tg}\subset \mathcal{S}_{tg}$ enclosed by $\gamma$.  
As before, we subdivide the rectangular surface $S_{tg}$ into $N\times N$ elementary plaquettes, with associated elementary
loops $\tilde{\gamma}_{ij}$ based at $p$ and their holonomies $\tilde{W}_{ij}$ (cf. eqs.\ (\ref{gammatilde}) and
(\ref{wtilde}) above). Because of the abelian nature of the holonomies, we can multiply them together in any order to obtain
the holonomy of $\gamma$,
\begin{equation}
W_\gamma=\prod_{i,j=0}^{N-1} \tilde{W}_{ij}.
\label{together}
\end{equation}
The total rotation angle $\alpha_\gamma$, associated to $W_\gamma$ according to eq.\ (\ref{rotmat}),
is simply given by adding the individual angles,
\begin{equation}
\alpha_\gamma =\sum_{i,j =0}^{N-1}\alpha_{\tilde{\gamma}_{ij}}\!\! \mod 2\pi,
\label{addangles}
\end{equation} 
which is an exact relation independent of $N$. The angles $\alpha_{\tilde{\gamma}_{ij}}$ in eq.\ (\ref{addangles}) can
be related easily to the corresponding angles $\alpha_{\gamma_{ij}}$ of the plaquette loops $\gamma_{ij}$. 
Since the three-dimensional metric can be made block-diagonal, and since the surface $S_{tg}$ has the
topology of a disc, the two-dimensional metric of the surface can without loss of generality be chosen conformally
flat, i.e.\ of the form e$^\phi \delta_{ab}$, where $\delta_{ab}$ is the two-dimensional Euclidean metric and $\phi (x_1,x_2)$
a conformal factor. For any such choice, the holonomy of any loop in $S_{tg}$ will have the canonical form (\ref{rotmat}).
Moreover, since for a given $(i,j)$ the holonomies $W_{ij}$ and $\tilde{W}_{ij}$ are related by conjugation,
the corresponding rotation angles
$\alpha_{\gamma_{ij}}$ and $\alpha_{\tilde{\gamma}_{ij}}$ must be equal up to a sign. However, since we are working
with frames that all have the same orientation (relative to the normal vector $\hat n$) along $S_{tg}$, those two angles 
must be identical, and therefore the same must be true for their corresponding holonomy matrices, $\tilde{W}_{ij}=W_{ij}$. 
We thus obtain
\begin{equation}
\alpha_\gamma =\sum_{i,j =0}^{N-1}\alpha_{\gamma_{ij}}\!\! \mod 2\pi.
\label{addangles2}
\end{equation}

Our next task will be to express the rotation angle $\alpha_{\gamma_{ij}}$ for a small plaquette loop $\gamma_{ij}$
as a function of the Riemann curvature tensor at its base point $x_{ij}$. Let us first derive this relation for an arbitrary infinitesimal 
rectangular loop at $x_{ij}$ and then return to the case at hand. We will use a variant of eq.\ (\ref{plaqhol}) that is slightly more
convenient for our purposes.  
Consider a right-handed, orthonormal basis $(\hat{e}_1,\hat{e}_2,\hat{n})$ of the tangent
space $T_{x_{ij}}\cal{M}$, where $\hat{e}_1$ and $\hat{e}_2$ are tangent to the surface $S_{tg}$. Pick two linearly independent vectors 
$u$ and $v$ in the span of $\hat{e}_1$ and $\hat{e}_2$ or, equivalently, the span of $\dot{\Phi}$ and $\Phi'$ at $x_{ij}$ 
such that the pair $(u,v)$ is positively oriented, i.e. $u^1 v^2-u^2 v^1>0$. For small $\varepsilon>0$, consider the small loop 
$\gamma_\varepsilon\subset S_{tg}$ of side length $\varepsilon$ 
spanned by the vectors $\varepsilon u$ and $\varepsilon v$, which is defined as the image under $\Phi$ of the small 
parallelogram spanned by the pullbacks of these vectors. Assume its orientation is counterclockwise, starting from $x_{ij}$ in
the direction of $u$. 
To lowest nontrivial order in $\varepsilon$, its holonomy is then given by
\begin{equation}
W_{\gamma_\varepsilon}=\mathbb{1}- 
\varepsilon^2 R(u,v)+o(\varepsilon^2)=\mathbb{1}- 
\varepsilon^2 R(x_{ij})^{\cdot}{}_{\cdot\mu\nu}u^\mu v^\nu+o(\varepsilon^2).
\label{paraexp}
\end{equation}  
Since we have chosen an orthonormal tangent space basis at $x_{ij}$, $W_{\gamma_\varepsilon}$ is an (infinitesimal) rotation in 
standard form (\ref{rotmat}), which for a small angle $\alpha_\gamma$ can be expanded as
\begin{equation}
W_\gamma =\left(\begin{array}{ccc}1 & 0 & 0 \\
0 & 1 & 0 \\
0 & 0 & 1\end{array}\right)+
\left(\begin{array}{ccc} 0 & -\alpha_\gamma & 0 \\
\alpha_\gamma & 0 & 0 \\0 & 0 & 0\end{array}\right)+ {\cal O}(\alpha_\gamma^2).
\label{rotmatexp}
\end{equation}
In order to extract the small angle $\alpha_{\gamma_\varepsilon}$ associated with the loop holonomy (\ref{paraexp}),
we consider the parallel transport of the vector $u\in T_{x_{ij}}{\cal M}$ around the loop spanned by $\varepsilon u$ and $\varepsilon v$,
and project the result onto $v\in T_{x_{ij}}{\cal M}$. From eq.\ (\ref{paraexp}), one finds
\begin{equation}
\langle v, W_{\gamma_\varepsilon} u\rangle =\langle v,u\rangle -\varepsilon^2 \langle v,R(u,v) u \rangle +\dots,
\label{vproj1}
\end{equation}
while evaluating the same expression in the given basis and using (\ref{rotmatexp}) gives
\begin{equation}
\langle v, W_{\gamma_\varepsilon} u\rangle =u^1 v^1+u^2 v^2+\alpha_{\gamma_\varepsilon}(u^1 v^2-u^2 v^1)+\dots \, .
\label{vproj2}
\end{equation}
The first two terms in (\ref{vproj2}) are simply the scalar product $\langle v,u\rangle$ in the given orthonormal basis, and
the term multiplying the angle $\alpha_{\gamma_\varepsilon}$ is the area in tangent space of the flat parallelogram 
spanned by $u$ and $v$. It follows that the small-$\alpha_{\gamma_\varepsilon}$ expansion (\ref{vproj2}) can be written
in an invariant way, which does not depend on the choice of basis in the tangent space, namely,
\begin{equation}
\langle v, W_{\gamma_\varepsilon} u\rangle =\langle v,u\rangle+
\alpha_{\gamma_\varepsilon}\sqrt{\langle u,u\rangle \langle v,v\rangle -\langle u,v\rangle^2}+ {\cal O}(\alpha_{\gamma_\varepsilon}^2).
\label{vprojinv}
\end{equation} 

Taking into account that the contraction of the Riemann tensor appearing in eq.\ (\ref{vproj1})
is related to the so-called sectional curvature $K_s(u,v)$ of the surface $S_{tg}$ via
\begin{equation}
K_s(u,v)=-\frac{\langle v,R(u,v) u \rangle}{\langle u,u\rangle \langle v,v\rangle -\langle u,v\rangle^2}
\label{seccurv}
\end{equation}
(see e.g. \cite{heckman}), we can to lowest order in $\varepsilon$ express the rotation angle $\alpha_{\gamma_\varepsilon}$ 
as a function of the sectional curvature by combining relations (\ref{vproj1}) and (\ref{vproj2}), yielding
\begin{equation}
\alpha_{\gamma_\varepsilon}= \varepsilon^2 \sqrt{\langle u,u\rangle \langle v,v\rangle -\langle u,v\rangle^2}\, K_s(u,v)+o(\varepsilon^2).
\label{alphaeps}
\end{equation}
Recall that the sectional curvature $K_s(u,v)$ depends on the two-dimensional plane in $T_{x_{ij}}{\cal M}$ spanned by 
the vectors $u$ and $v$, but not on the specific choice of the spanning vectors. Moreover, for the special case that 
the plane is tangent to a totally geodesic surface $S_{tg}$, as we are currently considering, the sectional curvature coincides with
the Gaussian curvature $K_G(x_{ij})$ of $S_{tg}$ at this point. This reflects the fact that a totally geodesic surface only carries intrinsic
curvature, but is not curved extrinsically with respect to the embedding manifold $({\cal M},g_{\mu\nu})$. Note furthermore that
the prefactor of the sectional curvature $K_s$ on the right-hand side of eq.\ (\ref{alphaeps}) is the area of a small parallelogram
spanned by the vectors $\varepsilon u$ and $\varepsilon v$. To leading order in $\varepsilon$, this {\it coincides} with the 
area $A(\gamma_\varepsilon)$ enclosed by the small rectangular loop $\gamma_\varepsilon$ in $S_{tg}$. Because 
of the linear dependence of (\ref{alphaeps})
on the lengths of the two vectors $u$ and $v$, if we had chosen a parallelogram with unequal
sides $\varepsilon_1$ and $\varepsilon_2$, the factor $\varepsilon^2$ would simply have been replaced by the product 
$\varepsilon_1 \varepsilon_2$.

We therefore have arrived at a straightforward geometric expression for the infinitesimal rotation angle $\alpha_{\gamma_\varepsilon}$ 
associated with the parallel transport
around an infinitesimal rectangular loop $\gamma_\varepsilon$ 
in a totally geodesic surface $S_{tg}$ embedded in a three-dimensional Riemannian manifold.
The angle is given simply by the value of the sectional curvature at the base point of $\gamma_\varepsilon$, 
multiplied by the area enclosed by the loop,
\begin{equation}
\alpha_{\gamma_\varepsilon}= A(\gamma_\varepsilon) K_s(u,v)+o(\varepsilon^2).
\label{alphaarea}
\end{equation}
Note that unlike the Wilson loop (\ref{trwg}), obtained by taking the trace of the holonomy, formula (\ref{alphaarea}) is sensitive
to the sign of the infinitesimal rotation angle. 
If the Gaussian curvature of $S_{tg}$ at the point $x_{ij}$ is positive (like that of a two-sphere, say), 
the rotation angle is also positive. Conversely, if the surface has negative Gaussian curvature, which means that the geometry
around $x_{ij}$ resembles a saddle point, the rotation angle will be negative.

We can now return to our tiling of a macroscopic piece $S_{tg}$ of a totally geodesic surface, enclosed by a rectangular loop
$\gamma$. Approximating the rotation angle $\alpha_{\gamma_{ij}}$ of each of the $N^2$ oriented plaquette loops in the 
sum (\ref{addangles2}) by the leading-order contribution of eq.\ (\ref{alphaarea}), and taking the limit $N\rightarrow\infty$, as
we did in Sec.\ \ref{wilsonloopgrav} when deriving the nonabelian Stokes' theorem.
This leads to a continuum expression for the total rotation angle in terms of an area integral, namely,
\begin{equation}
\alpha_\gamma=\int_{S_{tg}}\!\! K_s(u,v) dA \mod 2\pi,
\label{preformula}
\end{equation} 
where $u$ and $v$ refers to an arbitrary choice of a pair of smooth and everywhere linearly independent tangent vector 
fields. While for simplicity we have focused on a rectangular surface, it is clear that an embedded surface of any shape and with 
a suitably regular boundary can be 
decomposed into small rectangular tiles like the ones used above, possibly up to smoothing out some corners along 
boundaries.\footnote{We do not consider curves with cusps.} This will affect neither the 
loop nor the area integrations, so that eq.\ (\ref{preformula}) will continue to hold.

As a final remark, since the integral (\ref{preformula}) is over a non-infinitesimal surface $S_{tg}$, note that the 
resulting angle $\alpha_\gamma$ will in general not be small, unless
the curvature $K_s$ satisfies appropriate bounds on $S_{tg}$. Because of the compact range of the angle, it is then no longer meaningful to distinguish between positive and negative curvature. This is of course a general feature of holonomies of
compact groups, independent of their abelian or nonabelian character. -- In the next section we will show 
when and how eq.\ \rf{preformula} can be generalised to an expression that is invariant under deformations
of the surface enclosed by the loop $\gamma$.

\subsection{Surface-independence and geometric flux in three dimensions}
\label{sec:threedim}

The presence of a totally geodesic surface ${\cal S}_{tg}$ enabled us in Sec.\ \ref{sec:d3} to establish a relation between the 
(oriented) rotation angle of the holonomy of an arbitrary, non-selfintersecting closed curve $\gamma$ contained in 
${\cal S}_{tg}$ and the total curvature of
the totally geodesic surface $S_{tg}\subset{\cal S}_{tg}$ enclosed by $\gamma$. Taking this as a starting point, we will now show
how one can generalise this to a more powerful relation, in the spirit of Stokes' theorem. For the construction to apply, we
need not just an isolated totally geodesic surface, but a one-parameter family, in the form of a foliation of $M$ into totally
geodesic surfaces. Our discussion will be local in nature and will therefore not address the question when such a foliation exists 
globally in a given Riemannian manifold.

An important example of a situation where such a foliation occurs is 
in the presence of a hypersurface-orthogonal Killing vector field $\xi^\mu$
on $\cal M$. It is not the most general case of a foliation into totally geodesic surfaces, but it has the advantage of being familiar
from many physics applications. We will present a surface-independent version of eq.\ \rf{preformula} for this case below.
The condition of hypersurface-orthogonality means that the distribution orthogonal
to the Killing vector $\xi^\mu$ (the collection of subspaces in $T_p{\cal M}$ orthogonal to $\xi$ for all points $p$) is integrable.
A necessary and sufficient condition for a Killing vector field $\xi$ to be hypersurface-orthogonal is 
\begin{equation}
\xi_{[\lambda} \nabla_\mu \xi_{\nu ]} 
= 0,
\label{frobeniustest}
\end{equation}
which is a version of Frobenius' theorem \cite{wald}. The square brackets in eq.\ (\ref{frobeniustest}) denote a total
antisymmetrization over the three indices. It is easy to see that the hypersurfaces orthogonal to the Killing vector are totally
geodesic. Consider a point $p$ in some two-dimensional hypersurface ${\cal S}\subset {\cal M}$ and a vector
$u_p\in T_p{\cal M}$ that is tangent to $\cal S$ and therefore satisfies $\langle \xi,u_p\rangle=0$ at $p$. On the one hand, 
the initial values $(p,u_p)$ determine
a geodesic $\tilde{\gamma}(\tau)$ in $\cal S$ with respect to the metric $\tilde g$ induced on $\cal S$ from $({\cal M},g)$. 
On the other hand, they
also determine a unique geodesic $\gamma(\tau)$ in $\cal M$, satisfying $\gamma(0)=p$ and $\dot{\gamma}(0)=u_p$.
However, because of
the Killing vector property, the scalar product $\langle \xi,\dot{\gamma}(\tau)\rangle$ of the Killing vector and the
tangent vector to the geodesic stays constant along $\gamma$. Since at the initial point we have
$\langle \xi,\dot{\gamma}(0)\rangle\equiv \langle \xi,u_p\rangle =0$, the geodesic's tangent vector is orthogonal
to the Killing vector everywhere along $\gamma$. This implies that the geodesic stays inside the hypersurface $\cal S$ and is
identical with the previous geodesic $\tilde{\gamma}$, thus proving that $\cal S$ is totally geodesic.  

\begin{figure}[t]
\centering
\includegraphics[width=0.4\textwidth]{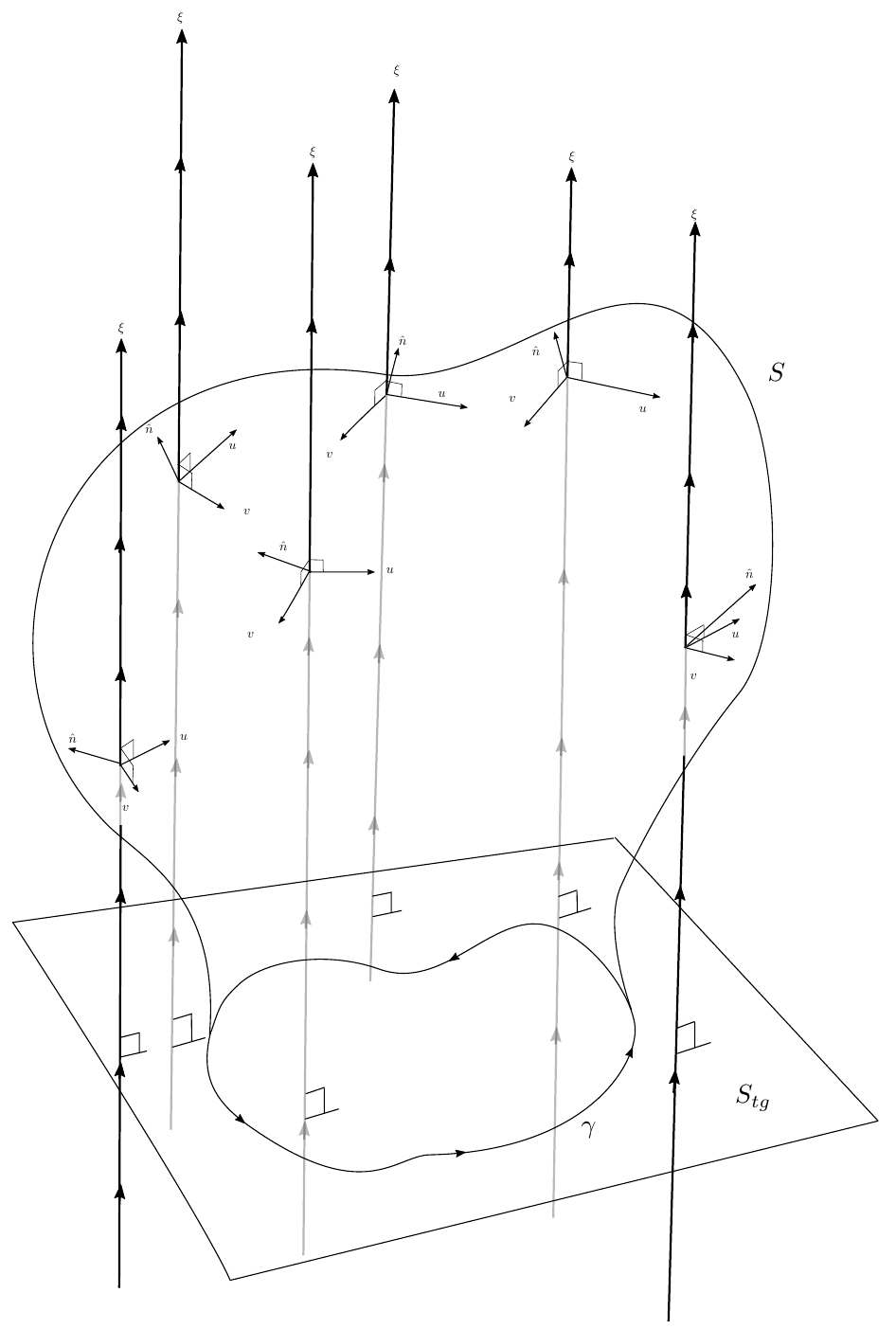}
\caption{\label{geometricflux} Quantities associated with the derivation of the geometric flux formula \eqref{surfint1}.
The only foliation leaf shown is the one containing the loop $\gamma$, whose holonomy matrix is characterised by the angle
$\alpha_\gamma$. The area integral that equals $\alpha_\gamma$ can be taken over the totally geodesic surface $S_{tg}$
in the leaf of $\gamma$ or any other smoothly deformed surface $S$, as along as its boundary remains fixed. The Killing vector
$\xi$ is everywhere perpendicular and the pair of vectors $(u,v)$ is everywhere tangent to the leaves of the foliation.}
\end{figure}

Let us assume the presence of a hypersurface-orthogonal Killing vector field $\xi$ and 
consider a surface $S_{tg}$ with the topology of a disc that lies
in one of the totally geodesic hypersurfaces, ${\cal S}_{tg}$, of the associated foliation of $\cal M$. Denote by $\gamma$ the loop that 
runs counterclockwise along the boundary $\partial S_{tg}$, with some arbitrarily chosen base point $p$, 
just like described in Sec.\ \ref{sec:d3}. We will generalise the relation \rf{preformula} such that it applies to any surface $S$ 
that can be obtained from $S_{tg}$ by a smooth deformation, while leaving the boundary fixed. The explicit expression is 
\begin{equation}
\alpha_\gamma := \int_S K_s(u,v)\, \frac{\xi^\mu \hat{n}_\mu}{|\xi|} \ d A  \mod 2\pi,
\label{surfint1}
\end{equation}
where $\hat{n}^\mu$ is the unit normal vector to the surface $S$, $dA$ is the invariant area element on $S$, and 
$|\xi |:=\sqrt{ \langle \xi,\xi \rangle} $ denotes
the norm of the vector $\xi$. The smooth vector fields
$u$ and $v$ are chosen arbitrarily, such that they span the space tangent to the leaves of the foliation and
therefore are orthogonal to the Killing vector $\xi$ everywhere. The new, deformed surface $S$ will in general not be
totally geodesic; only its boundary will remain in ${\cal S}_{tg}$. Nevertheless, the two-dimensional integration over $S$ on
the right-hand side of eq.\ (\ref{surfint1}) is well defined and moreover can be shown to be {\it independent of} $S$. 
Because the vectorial quantity projected onto the normal vector in the integrand depends only on the metric, 
we refer to the integrand of eq.\ \rf{surfint1} as a ``geometric flux" (see Fig.\ \ref{geometricflux} for illustration). 

For the special choice $S=S_{tg}$, expression (\ref{surfint1}) reduces to the previous relation \rf{preformula}, 
because the inner product $\langle \xi/|\xi|, \hat{n}\rangle$ of the normalized Killing vector and the normal vector $\hat{n}$
equals unity. Recall from Sec.\ \ref{sec:d3} that the normal vector $\hat n$ points in the 3-direction of a positively oriented
frame associated with the totally geodesic surface. Furthermore, we choose the direction of the Killing vector field such that
it is aligned with that of $\hat n$ on any of the totally geodesic leaves of the foliation. Whenever $S_{tg}$ is deformed into
another surface $S$, the normal vector $\hat n$ is deformed smoothly with it. Of course, at a general point of the deformed
surface $S$, its normal $\hat n$ will generally no longer be parallel to the Killing vector $\xi$.

To demonstrate the surface-independence of eq.\ (\ref{surfint1}), we invoke the standard Stokes' theorem for a three-dimensional compact oriented 
Riemannian manifold $N$ with boundary $\partial N$, which in Gauss law form reads
\begin{equation}
\int_N (\nabla_\mu X^\mu )\, dV =\int_{\partial N}(\hat{n}_\mu X^\mu ) \, dA
\label{stokes3d}
\end{equation}
for a vector field $X^\mu$ on $N$, where $\hat{n}^\mu$ is the outward-pointing unit normal vector on the boundary $\partial N$, and 
$dV$ and $dA$ denote the invariant volume elements on $N$ and $\partial N$ respectively. In the case at hand, the vector 
$X^\mu$ in eq.\ (\ref{stokes3d}) is given by
\begin{equation}
X^\mu= K_s(u,v)\, \frac{\xi^\mu }{|\xi|}, 
\label{xvec}
\end{equation}
and proving surface-independence is tantamount to showing that the divergence of the vector (\ref{xvec}) vanishes,
\begin{equation}
\nabla_\mu K_s(u,v) \frac{\xi^\mu}{|\xi|}\equiv
K_s(u,v)\nabla_\mu\frac{\xi^\mu}{|\xi|} +  \frac{\xi^\mu}{|\xi|}\nabla_\mu K_s(u,v) =0.
\label{divint}
\end{equation}
The first term on the right-hand side of eq.\ (\ref{divint}) vanishes by virtue of the Killing equation, since
\begin{equation}
\nabla_\mu\frac{\xi^\mu}{|\xi|}=
\left( \frac{1}{|\xi|}g^{\mu \nu} - \frac{1}{|\xi|^3}      \xi^\mu\xi^\nu \right)   \nabla_\mu\xi_\nu =
\left( \frac{1}{|\xi|}g^{\mu \nu} - \frac{1}{|\xi|^3}      \xi^\mu\xi^\nu \right)  \frac{1}{2}( \nabla_\mu\xi_\nu +\nabla_\nu\xi_\mu )= 0.
\label{killcalc1}
\end{equation}
The vanishing of the second term in eq.\ (\ref{divint}) is easiest to demonstrate in a particular coordinate system, which implies its
vanishing in general. The existence of a hypersurface-orthogonal Killing vector means that there exist local coordinates
adapted to the foliation in which the metric takes the block-diagonal form 
\begin{equation}
g_{\mu\nu}=\left(\begin{array}{ccc} E(x_1,x_2) & F(x_1,x_2) & 0 \\
F(x_1,x_2) & G(x_1,x_2) & 0 \\
0 & 0 & H(x_1,x_2)\end{array}\right)
\label{metrichyp}
\end{equation}
and the Killing vector is given by $\xi^\mu=(0,0,c)$, for some real constant $c$. Since the scalar quantity
$K_s(u,v)$ only depends on $x_1$ and $x_2$, one easily verifies that $\xi^\mu\nabla_\mu K_s\! =\! 0$, thus showing
that the expression (\ref{divint}) vanishes, as asserted earlier. We have therefore derived a continuity equation for the 
geometric flux of the vector quantity $K_s \xi^\mu/|\xi|$, which implies the surface-independence of eq.\ (\ref{surfint1}). 
Note that in the more general case where we have a foliation by totally geodesic surfaces, but no Killing vector, the metric 
can be brought into the same block-diagonal form (\ref{metrichyp}), but with $H(x_1,x_2)$ replaced by a more general function 
$H(x_1,x_2,x_3)$ \cite{cartan}. In this case we can still compute the angle $\alpha_\gamma$ by integrating over the surface
$S_{tg}$ according to eq.\ (\ref{preformula}), but there is no immediate analogue of the proof of surface-independence.

To summarize, for a three-dimensional Riemannian manifold that allows for a foliation by totally geodesic surfaces, we have
established a {\it new} relation, eq.\ (\ref{surfint1}), between the rotation angle of the holonomy of a macroscopic loop $\gamma$ 
lying in a leaf ${\cal S}_{tg}$ of the foliation and a surface integral over the sectional curvature of the surface $S\subset {\cal S}_{tg}$ 
with boundary $\gamma$. By virtue of the existence a conserved geometric flux, the surface integral is invariant under smooth
deformations of the surface $S$ away from the leaf ${\cal S}_{tg}$. In the next section, we will present a nontrivial example in
three dimensions, where both sides of the geometric flux formula (\ref{surfint1}) can be computed, including a surface integral over
a non-totally geodesic surface $S$.

\subsection{A three-dimensional example}
\label{sec:S3}

As an illustration of the general construction of the conserved geometric flux we will now discuss a nontrivial example,
where the underlying curved manifold is the three-sphere $S^3$ of constant positive curvature.
There are two convenient ways of parametrising points on the three-sphere, either in terms of coordinates
$(x_1,x_2,x_3,x_4)\in \mathbb{R}^4$ satisfying $x_1^2+x_2^2+x_3^2+x_4^2=r^2$, where the radius $r>0$ of the submanifold 
$S^3\subset \mathbb{R}^4$ sets the scale for the scalar curvature $R=6/r^2$, or in terms of three angular variables
$(\psi,\theta,\phi)$, which are related to the Cartesian coordinates $x_i$ by
\begin{equation}
x_1=r\cos\psi,\;\;
x_2=r\sin\psi\cos\theta,\;\;
x_3=r\sin\psi\sin\theta\cos\phi,\;\;
x_4=r\sin\psi\sin\theta\sin\phi,
\label{defcoord}
\end{equation} 
with ranges $0 \le \psi \le \pi$, $0 \le \theta \le \pi$ and $0 \le \phi < 2 \pi$.
In terms of the latter, the metric on $S^3$ reads 
\begin{equation}
g_{\mu \nu}(\psi,\theta,\phi) =
\begin{pmatrix} 
r^2 & 0 &0 \\
0 & r^2 \sin^2{\psi} &0  \\
0 & 0 &  r^2 \sin^2{\psi}  \sin^2{\theta}
\end{pmatrix},
\end{equation}
which exhibits the usual coordinate singularities at $\psi=0,\pi$ and $\theta=0,\pi$. 

The three-sphere is an example of a maximally symmetric space and has six Killing vectors fields $\xi^\mu$,
satisfying Killing's equation
\begin{equation}
\nabla_{(\mu} \xi_{\nu)} := \nabla_\mu \xi_\nu + \nabla_\nu \xi_\mu = 0.
\label{KE}
\end{equation}
They can be identified with the six generators of the global $SO(4)$-isometry, which in Cartesian coordinates take the familiar form 
$\xi_{i,j}=x_i\partial_j -x_j\partial_i$, $i\not= j$, for a rotation in the $i$-$j$ plane. Since these vector fields are tangent to spherical
shells in $\mathbb{R}^4$, they reduce on $S^3$ to sections of the tangent bundle $TS^3$. Their explicit form in
angular coordinates is given by   
\begin{eqnarray}
&& \xi_{1,2}^\mu = (\cos\theta,-\cot\psi \sin\theta, 0),\crcr
&& \xi_{1,3}^\mu = (\sin\theta\cos\phi,\cot\psi\cos\theta\cos\phi,-\cot\psi\sin\phi/\sin\theta),\crcr
&& \xi_{1,4}^\mu = (\sin\theta\sin\phi,\cot\psi\cos\theta\sin\phi,\cot\psi\cos\phi/\sin\theta), \crcr
&& \xi_{2,3}^\mu = (0,\cos\phi,-\cot\theta\sin\phi),\crcr
&& \xi_{2,4}^\mu = (0,\sin\phi,\cot\theta\cos\phi),\crcr
&& \xi_{3,4}^\mu = (0,0,1).
\label{KillingS3}
\end{eqnarray}

Taking into account the Christoffel symbols $\Gamma$, which in matrix notation 
$(\Gamma_\mu)^\lambda{}_\nu :=  \Gamma^\lambda_{\mu\nu}$ take the form 
\begin{eqnarray}
\Gamma_\theta && \!\!\!\!\!\!\!\!\!\!
=
\begin{pmatrix}
0 & - \cos \psi \sin \psi & 0\\
{\cot \psi } & 0 & 0 \\
0 & 0 & {\cot \theta} 
\end{pmatrix},\;\;
\Gamma_\psi
=
\begin{pmatrix}
0 & 0 & 0\\
0 & {\cot \psi} & 0 \\
0 & 0 & {\cot \psi }
\end{pmatrix}, \crcr
&&\;\; \Gamma_\phi
=
\begin{pmatrix}
0 & 0 & -\cos{\psi} \sin{\psi} \sin^2{\theta}\\
0 & 0 & -\cos{\theta} \sin{\theta} \\
{\cot \psi } & {\cot \theta} & 0
\end{pmatrix},
\label{chri1}
\end{eqnarray}
one easily verifies that the vectors \eqref{KillingS3} satisfy the Killing equations \eqref{KE}.

Also the nonvanishing components of the Riemann tensor $R^\kappa{}_{\lambda\mu\nu}$ of $S^3$ can
be written in a compact matrix form, with $({\mathsf{R}}_{\mu\nu})^\kappa{}_\lambda := R^\kappa{}_{\lambda\mu\nu}$
(not to be confused with the Ricci tensor $R_{\mu\nu}:= R^\kappa{}_{\mu\kappa\nu}$), namely,

\begin{eqnarray}
{\mathsf{R}}_{\psi\phi} && \!\!\!\!\!\!\!\!\!\!
=
\begin{pmatrix}
0 & 0 & \sin^2\psi \sin^2\theta\\
0 & 0 & 0\\
-1 & 0 &0
\end{pmatrix},\;\;
{\mathsf{R}}_{\psi\theta}
=
\begin{pmatrix}
0 & \sin^2\psi & 0\\
-1 & 0 & 0\\
0 & 0 &0
\end{pmatrix}, \crcr
&&\;\; {\mathsf{R}}_{\theta\phi}
=
\begin{pmatrix}
0 & 0 & 0\\
0 & 0 & \sin^2\theta \sin^2\psi\\
0 & -\sin^2 \psi &0
\end{pmatrix}.
\end{eqnarray}

The two-dimensional totally geodesic submanifolds of the round $S^3$ are ``equatorial" two-spheres, which 
in the embedding in $\mathbb{R}^4$ are intersections of three-dimensional planes through the origin with
the three-sphere.
An example are all points on the three-sphere satisfying $x_1=0$. They form a totally geodesic submanifold because
they are the fixed point set of an involutive isometry (i.e. an isometry that is its own inverse), in this case, the map
$(x_1,x_2,x_3,x_4)\mapsto (-x_1,x_2,x_3,x_4)$ on $S^3$ \cite{berger,klingenberg}.
However, due to symmetry, any plane obtained
from this one by an $SO(4)$-rotation is of course also a totally geodesic submanifold.  

To illustrate the geometric flux construction, we will without loss of generality consider a family of geodesic surfaces perpendicular
to the Killing vector field $\xi_{1,2}$ of the set \eqref{KillingS3}. Because the Killing vector vanishes at all points with
$\psi = \pi/2$ and $\theta =\pi/2$, we confine ourselves to a simply connected region of $S^3$ where $\xi_{1,2}$ does not vanish, and
its induced flow between the two surfaces depicted schematically in Fig.\ \ref{geometricflux} is well defined. Despite the fact that we therefore do not
have a {\it global} foliation of the manifold into totally geodesic surfaces, all necessary ingredients of our derivation in Sec.\ 3 are
present in this finite region, and give rise to a conserved flux. 

The loop $\gamma$ we consider for the construction is given by
\begin{equation}
\gamma(\lambda)\equiv (\psi (\lambda),\theta (\lambda),\phi (\lambda))=(\psi_0,\pi/2,\lambda),\;\; \lambda \in [0,2 \pi [,
\label{philoop}
\end{equation}
where the angle $\psi$ has been fixed to a value $\psi_0$ in the range $\pi/2 < \psi_0 <\pi$.
The curve \eqref{philoop} runs along the boundary of a totally geodesic disc $S_{tg}$ defined by
\begin{equation}
S_{tg}(\sigma,\tau)=(\sigma,\pi/2,\tau),\;\;\; \sigma\in [\psi_0,\pi],\;\; \tau\in [0,2 \pi [.
\label{stgdef}
\end{equation}
Comparing with the coordinate expressions \eqref{defcoord}, we see that all points of \eqref{stgdef} satisfy
$x_2=0$ and $S_{tg}$ therefore lies in a totally geodesic surface and is a totally geodesic disc as defined earlier.
Since the Killing vector $\xi_{1,2}^\mu$ reduces to $(0,-\cot\psi,0)$ on the disc $S_{tg}$, it is perpendicular to it everywhere.
(Note that $-\cot\psi$ is positive in the range considered.)
We choose the orientation of the unit normal vector $\hat{n}^\mu_{tg}$ to $S_{tg}$ to coincide with that of
the Killing vector, to yield
\begin{equation}
\hat{n}^\mu_{tg} := \xi^\mu_{1,2}/| \xi^\mu_{1,2} | = (0,\frac{1}{r\sin\psi},0)  .
\label{normaltg}
\end{equation}
on $ S_{tg}$. The orientation of the curve $\gamma$ in \eqref{philoop} has been chosen such that the normal $n_{tg}$, the tangent vector
$\dot{\gamma}$ and the inward-pointing normal vector to $\gamma$ tangent to $S_{tg}$ form a right-handed reference frame. 

We can verify the totally geodesic property by computing the second fundamental form $II(v,w)$ of eq.\ \eqref{sform} on $S_{tg}$. 
Since in three dimensions there is only a single normal vector, it suffices to establish that
\begin{equation}
\langle \hat{n}_{tg},\nabla_v w\rangle =0,\;\;\; \forall v,w\in TS_{tg},
\label{sectest}
\end{equation}
which is straightforward, using $v=(v^\psi,0,v^\theta)$, $w=(w^\psi,0,w^\theta)$, 
the Christoffel symbols \eqref{chri1} and the fact that $\theta=\pi/2$.

The induced metric on the totally geodesic surface in terms of the coordinates $(\psi,\phi)$ is given by
\begin{equation}
g_{tg}(\psi,\phi) = 
\begin{pmatrix} 
r^2 & 0  \\
0 & r^2 \sin^2 {\psi}  
\end{pmatrix}.
\end{equation}
The area $A_{tg}$ of the geodesic disc $S_{tg}$ is
\begin{equation}
A_{tg} 
= \int_0^{2 \pi}\!\!\! d \phi \int_{\psi_0}^{\pi}\! d \psi \; \sqrt{\det g_{tg}}
= 2 \pi r^2 (1 + \cos {\psi_0}).
\end{equation}
Noting that the sectional curvature $K_s$ is constant throughout $S^3$, $K_s=1/r^2$, the angle $\alpha_\gamma$ of 
eq.\ \eqref{preformula} is easily computed as
\begin{equation}
\alpha_\gamma
= \frac{1}{r^2} \ A_{tg}
=  2 \pi  (1+  \cos {\psi_0}).
\label{S3angletgs}
\end{equation}

Next, we will consider a non-totally geodesic surface $S_{ntg}$ that shares the boundary loop $\gamma (\lambda)$
of eq.\ \eqref{philoop}. It is given by
\begin{equation}
S_{ntg}(\sigma,\tau)=(\psi_0,\sigma,\tau), \;\;\; \sigma\in [\pi/2,\pi],\;\; \tau\in [0,2 \pi [,
\label{snontg}
\end{equation}
where $\psi_0$ was defined above in connection with eq.\ \eqref{philoop}. Like the surface $S_{tg}$, this disc is also
part of a two-sphere, but with a radius (in the embedding space) strictly smaller than $r$. This implies that its geodesics,
the great circles, are {\it not} geodesics of $S^3$. The unit normal vector $\hat{n}_{ntg}$ to $S_{ntg}$ is given by
\begin{equation}
\hat{n}_{ntg}^\mu=(-\frac{1}{r},0,0).
\label{normalntg}
\end{equation}
It is no longer collinear with the Killing vector $\xi^\mu_{1,2}$, but the two vector fields are still aligned in the sense
that their scalar product on $S_{ntg}$,  
\begin{equation}
\langle \hat{n}_{ntg},\xi_{1,2}\rangle = -r\cos\theta \geq 0,
\label{prodnxi}
\end{equation}
is positive inside the disc and vanishing along its boundary.

As a cross check, let us compute the second fundamental form on $S_{ntg}$. The analogue of eq.\ \eqref{sectest}
evaluates to 
\begin{equation}
\langle \hat{n}_{ntg},\nabla_v w\rangle = r\sin\psi_0\cos\psi_0 (v^\theta w^\theta+\sin^2\theta\, v^\phi w^\phi),
\label{secntg}
\end{equation}
which does not vanish for general tangent vectors $v,w\in TS_{ntg}$, confirming that the surface is not
totally geodesic. 
The induced metric on $S_{ntg}$ is
\begin{equation}
g_{ntg}(\theta, \phi)=
\begin{pmatrix} 
r^2\sin^2\psi_0 & 0   \\
0 & r^2 \sin^2 \psi_0\sin^2\theta
\end{pmatrix}.
\label{metntg}
\end{equation}

In terms of the ingredients needed for the arguments of Sec.\ 3 to apply, we still need to establish the existence of 
a smooth deformation of $S_{tg}$ to $S_{ntg}$, as illustrated in Fig.\ \ref{geometricflux}. We will not present an explicit map (which at any rate is highly 
nonunique), but argue that there are natural deformations, which are obtained by moving along the flow lines of the Killing
vector $\xi_{1,2}$. The first thing to note is that both discs, and the three-volume they enclose lie in a region away from
any zeros of $\xi_{1,2}$, where therefore a well-defined local foliation perpendicular to $\xi_{1,2}$ exists. On this region we could 
in principle introduce a local coordinate system adapted to the foliation to simplify calculations. However, in the case at hand this would not
lead to a further simplification because of the constancy of the sectional curvature. 

The flow of $\xi_{1,2}$ allows us to uniquely associate every point $p\in S_{ntg}$ with a point
$p_0\in S_{tg}$ that lies on the same orbit with regard to a rotation in the $x_1$-$x_2$ plane, restricted to $S^3$, generated by $\xi_{1,2}$.
For a given point $p=(\psi_0,\theta,\phi)\in S_{ntg}$, the angle $\omega (\theta,\phi)$ of the rotation by which it can be reached from
the corresponding point $p_0\in S_{tg}$ on the same orbit is given by
\begin{equation}
\omega (\theta,\phi)= \arctan (\cos\theta \tan\psi_0),
\label{angle12}
\end{equation}
which varies smoothly over the range $\omega\in [0,\pi -\psi_0]$ as a function of the angle $\theta\in [\pi/2,\pi]$, and is independent of
$\phi$. As one would expect, $d\omega/d\theta\rightarrow 0$ as one approaches the apex or centre of the spherical cap $S_{ntg}$,
$\theta\rightarrow\pi$. Since the rotation angle depends on $\theta$, one way of constructing a smooth deformation, parametrized by
some $\rho\in [0,1]$, would be by adjusting the rotation speed along each orbit, such that -- starting on $S_{tg}$ at $\rho=0$ --
all points on $S_{ntg}$ are reached simultaneously at $\rho=1$.  

Given the existence of a smooth deformation between $S_{tg}$ and $S_{ntg}$, our next task is to compute
the angle of formula \eqref{surfint1} by integrating over $S_{ntg}$. On this disc, the Killing vector 
$\xi_{1,2}^\mu = (\cos\theta,-\cot\psi_0 \sin\theta, 0)$ has the squared norm
\begin{equation}
| \xi_{1,2} |^2= r^2 (\cos^2\theta +\sin^2\theta \cos^2 \psi_0 ).
\label{normkvntg}
\end{equation}
Combining this with the volume element of the induced metric \eqref{metntg} and the scalar product \eqref{prodnxi}, one obtains
the angle $\alpha_\gamma$ of eq.\ \eqref{surfint1} as
\begin{eqnarray}
\alpha_\gamma &=& \int_0^{2\pi}\!\! d\phi \int_{\pi/2}^\pi d\theta \,\sqrt{\det g_{ntg}} \;
\frac{\langle \hat{n}_{ntg},\xi_{1,2}\rangle}{| \xi_{1,2} |}\, K_s(u,v)\crcr
&=& 
2\pi \sin^2 \psi_0 \int_{\pi/2}^\pi d\theta\; \frac{-\sin\theta\cos\theta}{\sqrt{\cos^2\theta+\sin^2\theta \cos^2\psi_0}}=
2\pi (1+\cos\psi_0),
\label{S3anglentg}
\end{eqnarray}
in agreement with our previous result \eqref{S3angletgs} for the analogous quantity on $S_{tg}$.

Lastly, let us compute the holonomy matrix $W_\gamma$ associated with the closed loop $\gamma (\lambda)$
of eq.\ \eqref{philoop} for comparison. Noting that the tangent vector is given by $\dot{\gamma}^\mu =(0,0,1)$,
one finds 
\begin{equation}
W_{\gamma}
= e^{- \int_{ \lambda = 0}^{2 \pi} d\lambda\, {\dot{\gamma}^\phi} \, \Gamma_\phi }
= e^{- {2 \pi} \Gamma_\phi }.
\label{holosphere}
\end{equation}
Since the matrix $\Gamma_\phi$ of \eqref{chri1} does not depend on the position along $\gamma$, 
there is no need for any path ordering in eq.\ \eqref{holosphere}. Explicitly, the holonomy matrix reads
\begin{equation}
W_{\gamma}=
\begin{pmatrix} 
\cos {(2 \pi \cos{\psi_0})} & 0 & \sin{(2 \pi \cos{\psi_0})} \sin{\psi_0}  \\
0 & 1 &0  \\
- {\sin{(2 \pi \cos{\psi_0})}  \over  \sin {\psi_0}} & 0 & \cos {(2 \pi \cos{\psi_0})} 
\end{pmatrix},
\label{WilsonloopS3}
\end{equation}
which shows that it is a rotation with angle 
\begin{equation}
\alpha_\gamma = 2 \pi \cos {\psi_0} \mod 2\pi,
\end{equation}
in agreement with the results \eqref{S3angletgs} and \eqref{S3anglentg}. 
The rotation matrix does not have standard form because the coordinate basis for the tangent space we are using is not orthonormal.  
As one would expect for parallel transport along a totally geodesic surface $S_{tg}$, the rotation only affects vectors 
in the tangent space $TS_{tg}$ to the surface, while leaving the normal direction invariant.

\subsection{Surface-independence and geometric flux in four dimensions}
\label{sec:4d}

As a further step, we will try to generalise the construction of Secs.\ \ref{sec:d3} and \ref{sec:threedim} to a four-dimensional
orientable Riemannian manifold $(\mathcal{M},g_{\mu\nu})$. We will assume the existence of two Killing vectors $\xi^\mu$ and
$\eta^\mu$ on $\mathcal{M}$, such that there exists a family of two-surfaces orthogonal to the group orbits generated by the
Killing vectors.\footnote{Such a group action is called ``orthogonally transitive", see \cite{stephani} for a detailed discussion
in a Lorentzian context.} This case is characterised by a pair of con\-di\-tions, namely,
\begin{equation}
(\nabla_{[\kappa}\xi_\lambda)\xi_\mu \eta_{\nu]}=0\;\; \land \;\; (\nabla_{[\kappa}\eta_\lambda)\eta_\mu \xi_{\nu]}=0.
\label{orthotrans}
\end{equation} 
A stronger condition that implies (\ref{orthotrans}) is the hypersurface-orthogonality of both Killing vectors,
\begin{equation}
\xi_{[\lambda}\nabla_\mu \xi_{\nu]}=0\;\; \land \;\; \eta_{[\lambda}\nabla_\mu \eta_{\nu]}=0,
\label{hyper2}
\end{equation}
which we have encountered earlier in the context of a single hypersurface-orthogonal Killing vector in three dimensions, 
eq.\ (\ref{frobeniustest}). In what follows, we will focus on the case where the two Killing vectors $\xi$ and $\eta$ do not commute, and
therefore generate a nonabelian group of isometries on $\mathcal{M}$.\footnote{The abelian case can be treated along the same
lines in a straightforward way.}
If eqs.\ (\ref{orthotrans}) are satisfied, one can introduce local coordinates $(x_1,x_2,x_3,x_4)$ in which the metric
takes the block-diagonal form
\begin{equation}
g_{\mu\nu}\! =\!\! \left(\begin{array}{cccc} 
{\rm e}^{M(\cdot)} & 0 & 0 & 0\\
0 & {\rm e}^{M(\cdot)}  & 0 & 0 \\
0 & 0 & W(x_1,x_2) {\rm e}^{\Psi (\cdot)} {\rm e}^{-2 x_4} & \Omega(x_1,x_2) W(x_1,x_2) {\rm e}^{\Psi(\cdot)} {\rm e}^{- x_4}\\
0 & 0 & \Omega(x_1,x_2) W(x_1,x_2) {\rm e}^{\Psi(\cdot)} {\rm e}^{- x_4} & W(x_1,x_2) ({\rm e}^{-\Psi(\cdot)}
+{\rm e}^{\Psi(\cdot)} \Omega(x_1,x_2)^2)
\end{array}\right)\! ,
\label{metrichyp2}
\end{equation}
where the functions $M$, $W$, $\Psi$ and $\Omega$ depend only on the first two coordinates $(x_1,x_2)$ and the function $W(x_1,x_2)>0$
is given by
\begin{equation}
W(x_1,x_2)^2=2\, {\rm e}^{2 x_4}\xi_{[\mu}\eta_{\nu]} \xi^\mu \eta^\nu,
\label{wdef}
\end{equation}
in terms of the two Killing vectors $\xi\! =\! \partial_3$ and $\eta\! =\! x_3\partial_3+\!\partial_4$. The ordered pair $(\xi,\eta)$ forms
a posi\-tively oriented zweibein in the $3$-$4$ plane.
Note that for hypersurface-ortho\-gonal
Killing vectors satisfying (\ref{hyper2}) we can set $\Omega(x_1,x_2)=0$, so that the metric (\ref{metrichyp2}) becomes diagonal.

It is straightforward to see that the
two-surfaces orthogonal to the group orbits spanned by the Killing vectors are totally geodesic, by the same argument
we used in the three-dimensional case. Consider a point $p$ in such a surface ${\cal S}\subset {\cal M}$ and a vector 
$u_p\in T_p{\cal M}$ that is tangent to $\cal S$ and therefore satisfies $\langle \xi,u_p\rangle =0$ and
$\langle \eta,u_p\rangle =0$ at $p$. Then the geodesic $\gamma$ with velocity vector $u_p$ starting at $p$ must remain in $\cal S$,
because the scalar product of its tangent vector with each of the two Killing vectors stays constant along $\gamma$.  
In the adapted coordinates $(x_1,x_2,x_3,x_4)$, each totally geodesic surface $\cal S$ is parametrized by the two
coordinates $x_1$ and $x_2$.

Let us now consider a closed, non-selfintersecting contractible curve $\gamma(\tau)$ 
lying in one of these totally geodesic surfaces ${\cal S}_{tg}\subset {\cal M}$, with base point $p$. 
Following our earlier treatment in Sec.\ \ref{sec:d3}, we would like to relate the holonomy of $\gamma$ to a curvature integral
over the totally geodesic disc $S_{tg}\subset {\cal S}_{tg}$ enclosed by $\gamma$.
 
To compute the sectional curvature $K_s(u,v)$ of the surface $S_{tg}$, one makes an arbitrary choice of a pair $(u,v)$ of linearly 
independent tangent vectors to $S_{tg}$ and uses formula (\ref{seccurv}), yielding
\begin{equation}
K_s(u,v)= -\frac{1}{2}\, {\rm e}^{-M(x_1,x_2)}(\partial_1^2+\partial_2^2)M(x_1,x_2).
\label{seccurv4d}
\end{equation}
As we argued at the beginning of Sec.\ \ref{tgsurfaces}, parallel transport around the curve $\gamma$ will never mix tangent vectors
in $TS_{tg}$ with normal vectors in $NS_{tg}$, which for a four-dimensional manifold $\cal M$ implies that the holonomy matrix $W_\gamma$
takes values in the abelian product group
\begin{equation}
W_\gamma \in SO(2) \times SO(2).
\label{subgroup4}
\end{equation}
We will call $\alpha_\gamma^\parallel$ the rotation angle associated with the tangent directions ($x_1$ and $x_2$) and 
$\alpha_\gamma^\perp$ the rotation angle associated with the normal directions ($x_3$ and $x_4$). 
In complete analogy with the corresponding construction in three
dimensions, eq.\ (\ref{preformula}), we find for the former 
\begin{equation}
\alpha_\gamma^\parallel=\int_{S_{tg}}\!\! K_s(u,v)\, dA \mod 2\pi.
\label{anglepar}
\end{equation} 
To determine the angle $\alpha_\gamma^\perp$, we use the same decomposition of the disc 
$S_{tg}$ into infinitesimal area elements and their associated plaquette loops. The small angle $\alpha_{\gamma_\varepsilon}^\perp$ 
associated with the holonomy $W_{\gamma_\varepsilon}$ of a small loop $\gamma_\varepsilon \subset S_{tg}$ of side length
$\varepsilon$ spanned by (suitably rescaled versions of) the tangent vectors $u$ and $v$ can be extracted from the action of the holonomy 
on the Killing vectors $\xi$ and $\eta$, which span the normal spaces $NS_{tg}$. The analogue of relation (\ref{vproj1}) one
needs to consider is
\begin{equation}
\langle \eta, W_{\gamma_\varepsilon}\xi\rangle =\langle \eta,\xi \rangle -\varepsilon^2\langle \eta,R(u,v)\xi\rangle +\dots\, .
\label{xiproj1}
\end{equation}
Note that the holonomy matrix $W_{\gamma_\varepsilon}$, when restricted to the $x_3$- and $x_4$-directions,
does not have the standard form of a rotation matrix, because the
metric (\ref{metrichyp2}) is not orthogonal in these directions. Instead, it will have the form
\begin{equation}
W_{\gamma_\varepsilon}|_{3,4}=G\Big( \mathbb{1}+
\left(\begin{array}{cc} 0 & -\alpha_{\gamma_\varepsilon}^\perp \\
\alpha_{\gamma_\varepsilon}^\perp & 0  \end{array}\right)+ {\cal O}((\alpha_{\gamma_\varepsilon}^\perp)^2)\Big) G^{-1}
= \mathbb{1} +G  \left(\begin{array}{cc} 0 & -\alpha_{\gamma_\varepsilon}^\perp \\
\alpha_{\gamma_\varepsilon}^\perp & 0  \end{array}\right) G^{-1}+  {\cal O}((\alpha_{\gamma_\varepsilon}^\perp)^2)   ,
\label{matrixconj}
\end{equation}
for some matrix $G\in GL(2,\mathbb{R})$, where the conjugation with $G$ of the standard $SO(2)$-generator will in
general produce a matrix with nonvanishing entries on the diagonal. However, this does not lead to any additional
complications when extracting the rotation angle, which appears in an invariant relation analogous to eq.\ (\ref{vprojinv}),
\begin{equation}
\langle \eta, W_{\gamma_\varepsilon} \xi\rangle =\langle \eta,\xi \rangle+
\alpha_{\gamma_\varepsilon}^\perp
\sqrt{\langle \xi,\xi \rangle \langle \eta,\eta \rangle  - \langle \xi,\eta \rangle^2}+ {\cal O}( (\alpha_{\gamma_\varepsilon}^\perp)^2).
\label{etaproj}
\end{equation}
We define a new function $K_\perp(u,v)$ by
\begin{equation}
K_\perp(u,v):=- \frac{  \langle \eta , R(u,v)\xi \rangle }{\sqrt{\langle u,u\rangle \langle v,v\rangle - \langle u,v \rangle^2}
\sqrt{\langle \xi,\xi \rangle \langle \eta ,\eta \rangle - \langle \xi ,\eta \rangle^2}}, 
\label{Kperp}
\end{equation}
where we have suppressed the dependence on the two Killing vectors in the notation. Like the corresponding expression
(\ref{seccurv}) for the sectional curvature $K_s(u,v)$, also $K_\perp(u,v)$ does not depend on the particular vectors $u$ and $v$, as
long as they span the tangent space $TS_{tg}$.
By using the new quantity $K_\perp(u,v)$ and combining eqs.\ (\ref{xiproj1}) and (\ref{etaproj}), we can rewrite eq.\ (\ref{etaproj}) as 
\begin{equation}
\alpha_{\gamma_\varepsilon}^\perp= \varepsilon^2 \sqrt{\langle u,u \rangle \langle v,v \rangle -
\langle u,v \rangle^2}\, K_\perp (u,v)+o(\varepsilon^2).
\label{alphaperpeps}
\end{equation}
From the metric (\ref{metrichyp2}), one computes 
\begin{equation}
K_\perp(u,v)= -\frac{1}{2}\, {\rm e}^{-M(x_1,x_2)}{\rm e}^{\Psi(x_1,x_2)} 
\big( (\partial_1\Omega(x_1,x_2))(\partial_2\Psi(x_1,x_2))-(\partial_2\Omega(x_1,x_2))(\partial_1\Psi(x_1,x_2))\big).
\label{kperpexp}
\end{equation}
Integrating up all infinitesimal angle contributions (\ref{alphaperpeps}), the rotation angle for the normal directions associated 
with the finite holonomy $W_\gamma$ is given by the surface integral
\begin{equation}
\alpha_\gamma^\perp=\int_{S_{tg}}\!\! K_\perp(u,v)\, dA \mod 2\pi.
\label{angleperp}
\end{equation} 
Note that the two rotation angles $\alpha_\gamma^\parallel$ and $\alpha_\gamma^\perp$ are not related. In particular, when
the Killing vector fields $\xi$ and $\eta$ are individually hypersurface-orthogonal, we have $\Omega=0$ and therefore any vector
normal to the geodesic surface $S_{tg}$ is unchanged after parallel-transporting it along the loop $\gamma$.

Similar to what we did in three dimensions, also in four dimensions we can show that the surface integral (\ref{anglepar}) for the rotation angle 
$\alpha_\gamma^\parallel$ for the tangential directions does not change under a smooth deformation of the integration surface away from 
the totally geodesic surface ${\cal S}_{tg}$, as long as the boundary of the disc $S_{tg}$ remains fixed. One only needs to recognise that the
integral (\ref{anglepar}) can be viewed as the right-hand side of the Stokes' theorem
\begin{equation}
\int_N d\omega =\int_{\partial N} \omega,
\label{stokes2}
\end{equation}
involving a three-dimensional compact oriented Riemannian manifold $N$ with boundary $\partial N$, 
a two-form $\omega$ and its exterior derivative, the three-form $d\omega$.
Eq.\ (\ref{stokes2}) is equivalent to eq.\ (\ref{stokes3d}) above, but written in terms of differential forms.

In the case at hand, the two-form $\omega$ is most conveniently expressed in terms of the Hodge dual $\star B$ of the two-form
$B$ on the {\it four}-dimensional manifold $\cal M$, which is defined by 
\begin{equation}
B:=\frac{\xi_\mu \eta_\nu }{\sqrt{\langle\xi ,\xi \rangle \langle\eta ,\eta \rangle-\langle\xi,\eta\rangle^2}}\, K_s(u,v)\, dx^\mu\wedge dx^\nu 
\label{Bmunu4d}
\end{equation} 
in terms of the two Killing vectors and the sectional curvature of $S_{tg}$. The corresponding dual two-form $\star B$ is given by
\begin{equation}
\star B=\frac{1}{2}\,\epsilon_{\kappa\lambda\mu\nu}\, \xi^\kappa \eta^\lambda \,
\frac{K_s (u,v)}{\sqrt{\langle\xi ,\xi \rangle \langle\eta ,\eta \rangle-\langle\xi,\eta\rangle^2}}\, dx^\mu\wedge dx^\nu,
\label{bstar}
\end{equation}
where the totally antisymmetric coefficients $\epsilon_{\kappa\lambda\mu\nu}$ are the components of the four-form
\begin{equation}
\epsilon=\sqrt{g}\, \epsilon_{\kappa\lambda\mu\nu}\, dx^\kappa \wedge dx^\lambda\wedge dx^\mu\wedge dx^\nu,
\label{4form}
\end{equation}
which is the natural volume element on $({\cal M},g_{\mu\nu})$. It is straightforward to see that the vanishing of the exterior derivative 
of a dual two-form, $d\star B=0$, is equivalent to the vanishing of the divergence 
\begin{equation}
\nabla_\mu B^{\mu\nu}=0,
\label{diverge}
\end{equation} 
expressed in terms of the components of the original two-form $B$. For the two-form (\ref{Bmunu4d}) at hand, eq.\ (\ref{diverge}) 
can be verified easily by explicit calculation in the particular coordinate system 
we have been working with, and is therefore satisfied in general. Having established that the two-form $d\star B$ vanishes on
the four-dimensional manifold $\cal M$ implies that its pullback on every embedded submanifold of $\cal M$ also vanishes. At the
same time, the two-form $\star B$ given by eq.\ (\ref{bstar}) also pulls back to any submanifold of $\cal M$. 

The relevance of these considerations for proving the invariance of the integral (\ref{anglepar}) under smooth surface deformations
comes from the fact that pulling back $\star B$ to the totally geodesic surface and integrating it over the disc $S_{tg}$ reproduces the integral on the
right-hand side of eq.\ (\ref{anglepar}). Imagine a smooth deformation of the surface $S_{tg}$ to some new surface $S$, obtained by moving each point
of $S_{tg}$ along one of a two-parameter family of flow lines to $S$ while keeping the boundary $\partial S_{tg}$ fixed. Assuming for simplicity
that $S$ and $S_{tg}$ only intersect along their common boundary, the flow lines sweep out a simply connected three-dimensional submanifold
$N$ of $\cal M$, with boundary $\partial N= S \cup S_{tg}$. We can then apply Stokes' theorem (\ref{stokes2}), with a vanishing left-hand side and
substituting $\omega =\star B|_{\partial N}$ on the right-hand side, thereby
proving the surface-independence of the computation of the angle $\alpha_\gamma^\parallel$, eq.\ (\ref{anglepar}).
In other words, we can write
\begin{equation}
\alpha_\gamma^\parallel=\int_{S} \star B \mod 2\pi, \;\; \;\;  ( \partial S =\gamma)
\label{angleparbbb}
\end{equation} 
with $\star B$ given by eq.\ (\ref{bstar}) and the understanding that $S$ is obtained from $S_{tg}$ through a smooth deformation. 
Like in the three-dimensional case, one can in principle identify a geometric flux through these two-dimensional surfaces, but it 
will depend on the choice of $N$ and is therefore a less immediate concept.

The same construction goes through for the rotation angle $\alpha_\gamma^\perp$ for the normal directions, 
using the function $K_\perp (u,v)$ instead of $K_s(u,v)$ in the
formulas. The counterpart of the two-form $B$ of eq.\ (\ref{Bmunu4d}) is the two-form
\begin{equation}
C:=\frac{\xi_\mu \eta_\nu }{\sqrt{\langle\xi ,\xi \rangle \langle\eta ,\eta \rangle-\langle\xi,\eta\rangle^2}}\, K_\perp (u,v)\, dx^\mu\wedge dx^\nu .
\label{C4d}
\end{equation} 
The proof of the vanishing of the divergence, $\nabla_\mu C^{\mu\nu}=0$, goes through because $K_\perp$ -- like $K_s$ -- depends only on the
coordinates $x_1$ and $x_2$, and not on $x_3$ and $x_4$. We conclude that the rotation angle can be obtained by integrating over any surface
$S$ that can be reached from the totally geodesic surface by a smooth deformation, yielding
\begin{equation}
\alpha_\gamma^\perp =\int_{S} \star C \mod 2\pi, \;\; \;\;  ( \partial S =\gamma)
\label{angleperpbbb}
\end{equation} 
in complete analogy with eq.\ (\ref{angleparbbb}). 

To summarize, we have considered a four-dimensional Riemannian manifold $({\cal M},g_{\mu\nu})$ with two noncommuting Killing vectors, 
generating group orbits that are perpendicular to a family of two-dimensional surfaces. 
We first showed that these surfaces are totally geodesic, and then considered the holonomy of a non-infinitesimal loop $\gamma \subset {\cal S}_{tg}$
contained in one of them. 
The holonomy $W_\gamma$ is characterised invariantly by two rotation angles, an angle $\alpha_\gamma^\parallel$ associated with a rotation 
of the tangent vectors in $T{\cal S}_{tg}$, and an angle $\alpha_\gamma^\perp$ associated 
with a rotation of the normal vectors in $N{\cal S}_{tg}$. The associated Wilson loop is given by
\begin{equation}
\mathrm{Tr}\,W_\gamma=2\cos(\alpha_\gamma^\parallel)+2\cos(\alpha_\gamma^\perp),
\label{4dWilsonexpl}
\end{equation}
which is seen to be a combined function of the two angles, and not sensitive to their signs. 
We demonstrated that both angles can be expressed as surface integrals of
two-forms, eqs.\ (\ref{anglepar}) and (\ref{angleperp}), depending on the Riemann curvature tensor via eqs.\ (\ref{seccurv}) and (\ref{Kperp})
respectively. Like in three dimensions, we were able to show that the integrations do not depend on the choice of surface spanning the loop $\gamma$.

\section{Summary and outlook}
\label{sec:conclusions}

Our investigation was prompted by the search for observables in nonperturbative quantum gravity that 
can capture information about the curvature of (quantum) spacetime. Because of a number of tantalising properties,
Wilson loops seem excellent candidates for such a role. Firstly, the holonomies on which they depend contain the complete information
on local curvature, as is clear from eq.\ (\ref{squareexp}). Secondly, gravitational Wilson loops are scalars with respect to diffeomorphisms (coordinate
transformations), although for the purpose of constructing quantum observables the loops on which they depend still need to be
specified appropriately. Thirdly, the loops are associated with length scales, like their intrinsic length or their diameter with respect to the ambient space. 
Determining the behaviour of suitable Wilson loop or holonomy operators as a function of scale in the quantum theory could allow us to interpolate between
a Planckian, short-scale regime and a large-scale regime, where one would look for evidence of a well-defined classical limit. The
blueprint for such an analysis is that of another scale-dependent operator, the spectral dimension, which 
in CDT quantum gravity was shown to exhibit a characteristic ``dynamical dimensional reduction" near the Planck scale, 
as well as a correct classical limit on larger scales \cite{spectral}.\footnote{see \cite{carlip} for a summary of related developments in other quantum gravity
approaches} Lastly, although holonomies on curved manifolds are usually impossible to compute in practice\footnote{This
can only be done for simple metrics $g_{\mu\nu}$ and simple paths $\gamma$; an explicit example we have come across in the literature is a computation
for circular loops on a Schwarzschild spacetime \cite{modanese,fredsted}.}(because of 
the need for path ordering),
this is not true for the piecewise flat geometries that appear in the path integrals of quantum gravity approaches based on simplicial manifolds,
like (C)DT and quantum Regge calculus.  
This comes from the fact that the triangular building blocks are flat in their interior, which means that nontrivial contributions
to a holonomy can only arise at points where the associated path crosses from one simplicial building block to an adjacent one. 
The biggest simplification occurs in the framework of Causal Dynamical Triangulations, 
because of the equilateral nature of its building blocks \cite{wilsoncdt}.     

To put these considerations on a firmer footing, we went in search of a relation between the holonomy of a macroscopic
loop and the curvature of a classical, Riemannian manifold, preferably of a simpler functional form than the already known nonabelian Stokes' theorem.  
For a class of curved manifolds with suitable isometries we were able to derive a set of new, exact relations that express the invariant 
angles characterizing the holonomy of a nonintersecting, contractible loop on a totally geodesic surface in three and four dimensions as
two-dimensional curvature integrals, eqs.\ (\ref{surfint1}), (\ref{anglepar}) and (\ref{angleperp}). We also demonstrated the existence of
a geometric flux in either dimension, which is responsible for the invariance of the area integrals under smooth surface deformations.
We have identified sufficient conditions on the Killing vectors for our construction to go through, eqs.\ (\ref{frobeniustest}) and (\ref{orthotrans}).
It would be interesting to investigate in more detail to what extent these conditions can be relaxed. 

Another potentially interesting generalization is to Lorentzian manifolds, whose holonomy 
groups are subgroups of the noncompact group $SO(1,d-1)$. Here, one will have to distinguish between 
several cases, depending on the signature of the induced metric
on the totally geodesic surfaces. At least in simple cases, the Riemannian analysis should essentially go through, with some of the compact abelian
subgroups substituted by their noncompact counterparts and some trigonometric functions substituted by hyperbolic ones.
Lightlike directions will require special attention, as is illustrated by an expression like eq.\ (\ref{surfint1}), which is no longer well defined when the Killing
vector $\xi$ is lightlike. 

Away from the spaces with symmetries we have studied, the nonabelian nature of the connection and the associated curvature tensor remains 
the main obstacle to establishing simple relations between holonomy and curvature. 
Although we have exploited the fact that in three and four dimensions holonomies along totally geodesic surfaces are abelian,
it should be emphasized that our setting is {\it not} purely abelian, in the sense that the holonomy of the manifold $({\cal M},g_{\mu\nu})$ is not required 
to be abelian, as illustrated by the example in Sec.\ \ref{sec:S3}, and also the holonomies lying on surfaces smoothly deformed away from the totally 
geodesic ones are not necessarily abelian. Note also that the curvature information captured by the invariant angles we have computed is not purely
two-dimensional. While the angle $\alpha_\gamma^\parallel$ of eqs.\ (\ref{surfint1}) and (\ref{anglepar})
is the averaged sectional curvature $K_s$ of the two-planes orthogonal to the Killing vector(s), the rotation angle 
$\alpha_\gamma^\perp$ of eq.\ (\ref{angleperp}), describing the effect of parallel transport on
the normal directions in four dimensions, involves also other components of the Riemann tensor, as
is clear from the explicit functional form of its integrand $K_\perp$ in eq.\ (\ref{kperpexp}).

Returning to our original motivation, the general lack of observables in nonperturbative quantum gravity, 
how can the insights gained in our investigation be used profitably in this context? 
Setting aside the question of how to average over loops to obtain well-defined quantum observables, 
several other issues still need to be addressed to make the construction meaningful in the quantum theory. 
For definiteness, let us consider what would be required in the case of CDT quantum gravity, where we already know that holonomies and 
Wilson loops can be defined and measured easily. The most important ingredient will be a suitable implementation on CDT's  piecewise flat 
manifolds of the concept of a totally geodesic surface. This should take into account that individual geodesics, although they
can be defined in fairly straightforward ways, e.g. as shortest curves with respect to (dual) link distance, are not very  
convenient to work with, since the curvature singularities of the triangulations lead to caustics and nonuniqueness 
already at the scale of the cutoff. More promising than some discretized version of a tensorial criterion involving the 
second fundamental form, eq.\ (\ref{sff}), may be a definition that involves surfaces perpendicular to one or more Killing vectors.
Although general simplicial spacetimes contributing to the path integral will not have any symmetries, it is plausible that
any ground state of geometry emerging from the quantum theory does possess such symmetries at a sufficiently
coarse-grained level. This should in turn be reflected in the existence of (approximate) Killing vectors, likewise at a
sufficiently coarse-grained length scale, a nontrivial concept that is currently being examined \cite{killing}.  

After defining a notion of totally geodesic surfaces for the simplicial geometries, a set of loops should be selected. A simple
choice would be to make them circular, invoking the geodesic link distance from a given centre point. The behaviour of the 
associated holonomies or Wilson loops could then be measured as a function of the radius of these circles. From this, it would
be interesting to understand
whether one can identify a length scale where the expectation values of the invariant angles become
small (relative to $\pi$), which could be a signature of quasi-classical behaviour. Whether suitable ``quantum holonomies" can
exhibit a behaviour where large Planckian fluctuations effectively average out and reproduce aspects of some
classical geometry remains at this stage an open question.
Answering it in the affirmative may require some ingenuity. Although it would be highly desirable,
there is of course no a priori guarantee that gravitational holonomies and Wilson loops can be used to understand the quasi-local
properties of gravity and spacetime in a nonperturbative regime. We hope to be able to report further progress on this issue
in the near future.


\subsection*{Acknowledgements}
We would like to thank Gert Heckman for pointing us to totally geodesic surfaces. -- This work was partly supported by the 
research program ``Quantum gravity and the search for quantum spacetime" of the Foundation for Fundamental Research on 
Matter (FOM, now defunct), financially supported by the Netherlands Organisation for Scientific Research (NWO).

\end{document}